\documentclass[a4paper, reprint, showkeys, nofootinbib, oneside,superscriptaddress]{revtex4-2}
\usepackage{graphicx}
\usepackage{lipsum} %ダミーテキストのやつ
\usepackage{color}
\usepackage[english]{babel}
\usepackage[utf8]{inputenc}
\usepackage[colorinlistoftodos, color=green!40, prependcaption]{todonotes}
\usepackage{amsthm}
\usepackage{mathtools}
\usepackage{physics}
\usepackage{xcolor}
\usepackage{graphicx}
\usepackage[left=23mm,right=13mm,top=35mm,columnsep=15pt]{geometry}
\usepackage{adjustbox}
\usepackage{placeins}
\usepackage[T1]{fontenc}
\usepackage{lipsum}
\usepackage{csquotes}

\usepackage{hyperref}
\usepackage{amsmath}
\usepackage{amsfonts}
\usepackage{amssymb}
\usepackage{enumitem}
\usepackage{float}
\usepackage{mathtools}
\usepackage{comment}
\usepackage{capt-of}
\usepackage{placeins} 
\usepackage{mathcomp}
\usepackage{natbib}

\makeatletter
\renewcommand{\eqref}[1]{(\ref{#1})}
\makeatother

\newcommand{\ctext}[1]{\raise0.2ex\hbox{\textcircled{\scriptsize{#1}}}}

\begin{document}

\title{Proof-of-Principle Experiment on a Displacement-Noise-Free Neutron Interferometer for Gravitational Wave Detection}

\author{Shoki Iwaguchi}
\affiliation{Department of Physics, Nagoya University, Furo-cho, Chikusa-ku, Nagoya, Aichi 464-8602, Japan}

\author{Takuhiro Fujiie}
\affiliation{Department of Physics, Rikkyo University, Ikebukuro Campus 3-34-1 Nishi-Ikebukuro, Toshima-ku, Tokyo 171-8501, Japan}

\author{Taro Nambu}
\affiliation{Department of Physics, Nagoya University, Furo-cho, Chikusa-ku, Nagoya, Aichi 464-8602, Japan}
\affiliation{RIKEN Center for Advanced Photonics, RIKEN, Hirosawa 2-1, Wako, Saitama 351-0198, Japan}

\author{Masaaki Kitaguchi}
\affiliation{The Kobayashi-Maskawa Institute for the Origin of Particles and the Universe, Nagoya University, Nagoya, Aichi 464-8602, Japan}
\affiliation{Department of Physics, Nagoya University, Furo-cho, Chikusa-ku, Nagoya, Aichi 464-8602, Japan}

\author{Yutaka Yamagata}
\affiliation{RIKEN Center for Advanced Photonics, RIKEN, Hirosawa 2-1, Wako, Saitama 351-0198, Japan}

\author{Kenji Mishima}
\affiliation{The Kobayashi-Maskawa Institute for the Origin of Particles and the Universe, Nagoya University, Nagoya, Aichi 464-8602, Japan}
\affiliation{High Energy Accelerator Research Organization, Tokai, Ibaraki 319-1106, Japan}

\author{Atsushi Nishizawa}
\affiliation{Department of Physics, Graduate School of Advanced Science and Engineering,
Hiroshima University, 1-3-1 Kagamiyama, Higashi-Hiroshima, Hiroshima 739-8526, Japan}
\affiliation{Astrophysical Science Center, Hiroshima University, Higashi-Hiroshima, Hiroshima 739-8526, Japan}
\affiliation{Research Center for the Early Universe (RESCEU), Graduate School of Science, The University of Tokyo, Bunkyo-ku, Tokyo 113-0033, Japan}

\author{Tomohiro Ishikawa}
\affiliation{Department of Physics, Nagoya University, Furo-cho, Chikusa-ku, Nagoya, Aichi 464-8602, Japan}

\author{Kenji Tsuji}
\affiliation{Department of Physics, Nagoya University, Furo-cho, Chikusa-ku, Nagoya, Aichi 464-8602, Japan}

\author{Kurumi Umemura}
\affiliation{Department of Physics, Nagoya University, Furo-cho, Chikusa-ku, Nagoya, Aichi 464-8602, Japan}

\author{Kazuhiro Kobayashi}
\affiliation{Equipment Development Support section, Technical Center of Nagoya University, Nagoya, Aichi 464-8602, Japan}

\author{Takafumi Onishi}
\affiliation{Equipment Development Support section, Technical Center of Nagoya University, Nagoya, Aichi 464-8602, Japan}

\author{Keiko Kokeyama}
\affiliation{Department of Physics, Nagoya University, Furo-cho, Chikusa-ku, Nagoya, Aichi 464-8602, Japan}
\affiliation{The Kobayashi-Maskawa Institute for the Origin of Particles and the Universe, Nagoya University, Nagoya, Aichi 464-8602, Japan}
\affiliation{School of Physics and Astronomy, Cardiff University, Cardiff CF24 3AA, United Kingdom}

\author{Hirohiko Shimizu}
\affiliation{Department of Physics, Nagoya University, Furo-cho, Chikusa-ku, Nagoya, Aichi 464-8602, Japan}

\author{Yuta Michimura}
\affiliation{Research Center for the Early Universe (RESCEU), Graduate School of Science, The University of Tokyo, Bunkyo-ku, Tokyo 113-0033, Japan}
\affiliation{Kavli Institute for the Physics and Mathematics of the Universe (Kavli IPMU), WPI, UTIAS, The University of Tokyo, Kashiwa, Chiba 277-8568, Japan}

\author{Seiji Kawamura}
\affiliation{Department of Physics, Nagoya University, Furo-cho, Chikusa-ku, Nagoya, Aichi 464-8602, Japan}

\begin{abstract}
\noindent
The displacement-noise-free interferometer (DFI) is designed to eliminate all displacement-induced noise while retaining sensitivity to gravitational wave (GW) signals. Ground-based DFIs suffer from physical arm-length limitations, resulting in poor sensitivity at frequencies below 1 kHz. To address this, previous research introduced a neutron-based DFI, which replaces laser light with neutrons and achieves exceptional sensitivity down to a few hertz. In this study, we conducted a proof-of-principle experiment using a pulsed neutron source at the Japan Proton Accelerator Research Complex (J-PARC). Despite practical constraints that led to deviations from the ideal experimental design, we optimized the setup and developed a novel analysis method that successfully cancels displacement noise while preserving simulated GW signals. This work presents the first successful demonstration of a neutron DFI and a neutron interferometer for GW detection.

\end{abstract}

\maketitle

\section{\label{sec:1}Introduction}

Since the first direct detections of gravitational waves (GW) — from binary black holes \cite{LIGO} and binary neutron stars \cite{Virgo} — over 90 GW events have been observed by the second-generation detector network comprising Advanced LIGO \cite{LIGO_ref} and Advanced Virgo \cite{Virgo_ref}, with KAGRA \cite{KAGRA_ref} joining subsequently.

Next-generation ground-based GW detectors, such as the Einstein Telescope (ET) \cite{ET} and the Cosmic Explorer (CE) \cite{CE}, are designed to observe GWs in the 10 Hz to 1 kHz range. At frequencies below this band, important scientific targets include intermediate-mass black holes \cite{IMBH_GW}, GWs associated with primordial black holes \cite{PBH_GW}, and primordial GWs \cite{PGW}. However, enhancing sensitivity at low frequencies is difficult due to displacement noise sources, including seismic and suspension thermal noise. A promising approach to suppress these noise sources is to use space-based detectors, such as the Laser Interferometer Space Antenna (LISA) \cite{LISA} and the Deci-hertz Interferometer Gravitational Wave Observatory (DECIGO) \cite{DECIGO_1, DECIGO_2}. Although space-based detectors can effectively eliminate suspension-related noise, their development demands substantial financial and temporal investment.

To mitigate displacement noise in ground-based interferometers, the displacement-noise-free interferometer (DFI) was proposed \cite{DFI_1, DFI_2}. It works by combining signals so that displacement noise components cancel out \cite{DFI_3}. This principle has been experimentally validated using specially designed interferometric configurations \cite{DFI_4, DFI_5}. The DFI's frequency sensitivity depends on the propagation time of laser light between mirrors. More precisely, a DFI with an arm length of $L$ has an effective frequency of $c/L$. Thus, even a DFI with an arm length of ~10 km is insufficiently sensitive below 1 kHz.

One method to solve this problem is the neutron DFI \cite{DFNI_Nishizawa}, which uses neutrons instead of laser light. Since neutron speed can be chosen within experimental constraints, the DFI's effective frequency can be reduced using neutrons with speeds slower than light. Another advantage of the neutron DFI is the simplification of the interferometer configuration. Previous studies have demonstrated that the number of optical components and the required neutron incidence angles can be reduced or optimized using multiple neutron speeds \cite{DFNI_Iwaguchi_1, DFNI_Iwaguchi_2}, with further refinements explored in \cite{DFNI_Kawasaki}.

In this study, we performed a proof-of-principle experiment to validate and assess the concept's feasibility. The experiment utilized a pulsed neutron source at the Materials and Life Science Experimental Facility (MLF) at the Japan Proton Accelerator Research Complex (J-PARC) \cite{J-PARC_1}.

This demonstration marks the first experimental realization of a neutron DFI and a neutron interferometer for GW detection. As noted in prior work \cite{DFNI_Nishizawa}, neutron-based detectors are expected to exhibit significantly greater sensitivity than ground-based laser interferometers at frequencies below a few Hz.

The paper is organized as follows: Section \ref{section:2} reviews the theoretical framework of the neutron DFI using a four-speed unidirectional configuration. Section \ref{section:3} compares the ideal and actual experimental setups. Section \ref{section:4} describes the analysis methodology and numerical simulations. Experimental results are presented in Section \ref{section:5}, and concluding remarks are given in Section \ref{section:6}.

\section{Review of Neutron DFI using four-speed unidirectional neutrons}
\label{section:2}

This section reviews the configuration of the interferometer examined in our proof-of-principle experiment. It also introduces the phasor diagram and the accompanying mathematical framework to demonstrate how displacement noise is canceled while GW signals are preserved. This framework serves as the basis for the analyses conducted throughout this paper.

\begin{figure}[H]
   \centering
   \includegraphics[clip,width=0.275\textwidth]{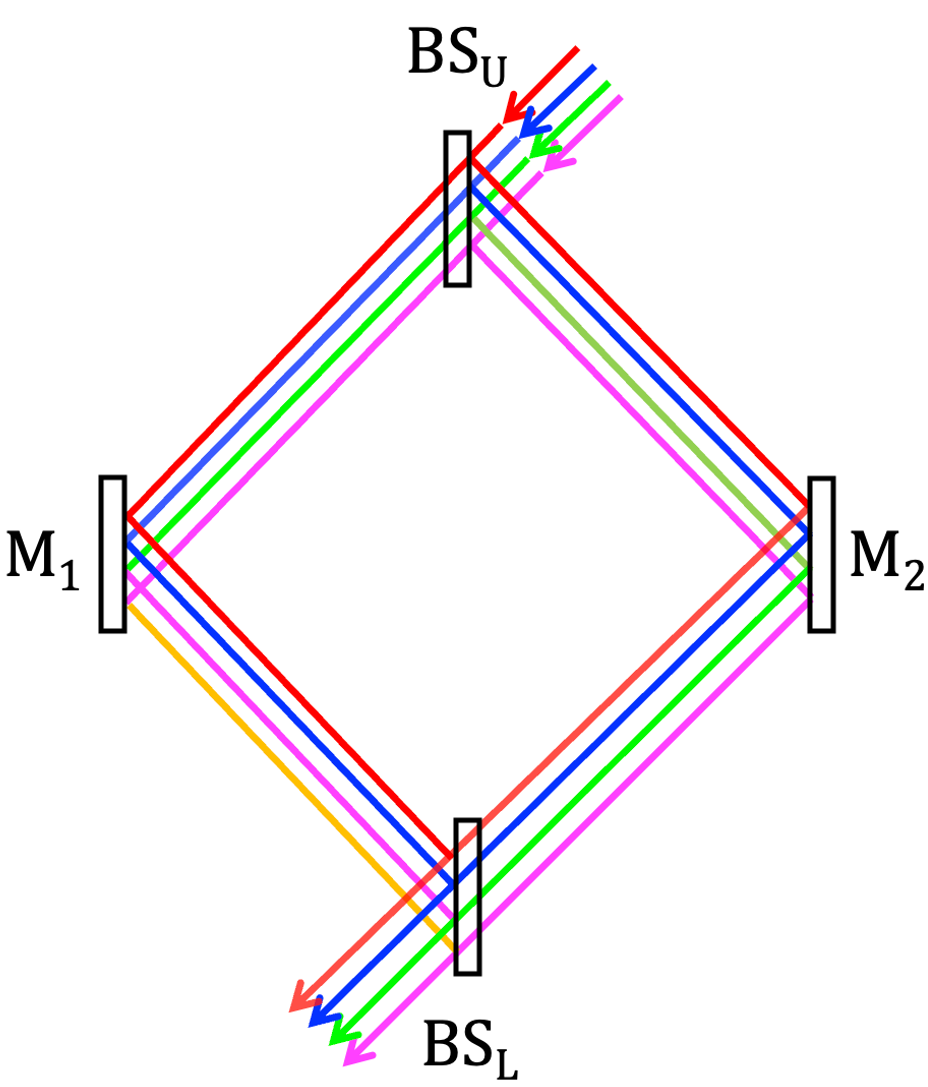}
   \caption{Mach-Zehnder interferometer illuminated with four-speed unidirectional neutrons. This interferometer comprises two beam splitters, $\mathrm{BS_U}$ and $\mathrm{BS_L}$, and two mirrors, $\mathrm{M_1}$ and $\mathrm{M_2}$. The red, blue, green, and magenta arrows indicate the trajectories of neutrons with the speeds of $v_\mathrm{1}$, $v_\mathrm{2}$, $v_\mathrm{3}$, $v_\mathrm{4}$, respectively.}
   \label{fig:1}
\end{figure}

The system under study is a neutron DFI, which utilizes unidirectional neutrons traveling at four distinct velocities $v_\mathrm{i}$, where $v_\mathrm{1} > v_{2} > v_{3} > v_{4}$. As shown in Fig. \ref{fig:1}, the configuration consists of a square Mach-Zehnder interferometer illuminated by neutrons at these four speeds. The interferometer comprises two beam splitters ($\mathrm{BS_U}$ and $\mathrm{BS_L}$) and mirrors ($\mathrm{M_1}$ and $\mathrm{M_2}$). The red, blue, green, and magenta arrows denote the trajectories of neutrons traveling at $v_{1}$, $v_{2}$, $v_{3}$, and $v_{4}$, respectively.

The phase shift of neutrons, $\phi_{\mathrm{l,i}}(t)$, induced by the displacement of optical elements or by a GW signal, is given by: $\phi_{\mathrm{l,i}}(t) = k_{\mathrm{i}} x_{\mathrm{l}}(t)$, where l denotes either a mirror ($\mathrm{M_1}$, $\mathrm{M_2}$), a beam splitter ($\mathrm{BS_U}$, $\mathrm{BS_L}$), or a simulated GW component ($\mathrm{GW_U}$, $\mathrm{GW_L}$), $x_l(t)$ is the time-dependent displacement of the corresponding element or signal, and $k_\mathrm{i}$ is the wave number associated with the neutron traveling at speed $v_\mathrm{i}$.

Let $T_{\mathrm{i}}$ represent the propagation time of a neutron along one side of the interferometer square. When neutrons at different speeds reach mirrors simultaneously, the associated phase shifts at $\mathrm{BS_U}$ and $\mathrm{BS_L}$ are given by $\phi_{\text{BSU},\mathrm{i}}(t - T_\mathrm{i})$ and $\phi_{\text{BSL},\mathrm{i}}(t + T_\mathrm{i})$, respectively. For GWs, we assume the neutron phase shift occurs along one leg of the interferometer path, around the midpoint between a mirror and a beam splitter (BS). This simplification yields GW-induced phase shifts of $\phi_{\text{GWU},\mathrm{i}}(t - T_i/2)$ and $\phi_{\text{GWL},\mathrm{i}}(t + T_\mathrm{i}/2$). Accordingly, the total phase shift for each neutron speed is: $\phi_{\mathrm{i}}(t) = \phi_{\mathrm{BS_U,\mathrm{i}}}(t - T_{\mathrm{i}}) + \phi_{\mathrm{GW_U,\mathrm{i}}}(t - T_{\mathrm{i}}/2) + \phi_{\mathrm{M_1,\mathrm{i}}}(t) + \phi_{\mathrm{M_2,i}}(t) + \phi_{\mathrm{GW_L,i}}(t + T_{\mathrm{i}}/2) + \phi_{\mathrm{BS_L,i}}(t + T_{\mathrm{i}})$, as shown in Table \ref{tab:1}.

\renewcommand\arraystretch{2.0}
\begin{table}[h]
  \centering
  \caption{Phase shifts caused by the displacement of mirrors, BSs, and the GW signal when unidirectional neutrons of different speeds hit the mirrors simultaneously.}
  \scalebox{0.65}{
  \begin{tabular}{c||c|c|c|c|c|c} \hline
    phase & $\mathrm{BS_U}$ & $\mathrm{GW_U}$ & $\mathrm{M_1}$ & $\mathrm{M_2}$ & $\mathrm{GW_L}$ & $\mathrm{BS_L}$ \\  \hline \hline
    $\phi_1(t)$ & $\phi_{\mathrm{BS_U},1} (t-T_1)$ & $\phi_{\mathrm{GW_U},1} (t-T_1/2)$ & $\phi_{\mathrm{M_1},1} (t)$ & $\phi_{\mathrm{M_2},1} (t)$ & $\phi_{\mathrm{GW_L},1} (t+T_1/2)$ & $\phi_{\mathrm{BS_L},1} (t+T_1)$ \\ \hline
    $\phi_2(t)$ & $\phi_{\mathrm{BS_U},2} (t-T_2)$ & $\phi_{\mathrm{GW_U},2} (t-T_2/2)$  & $\phi_{\mathrm{M_1},2} (t)$ & $\phi_{\mathrm{M_2},2} (t)$ & $\phi_{\mathrm{GW_L},2} (t+T_2/2)$  & $\phi_{\mathrm{BS_L},2} (t+T_2)$ \\ \hline
    $\phi_3(t)$ & $\phi_{\mathrm{BS_U},3} (t-T_3)$ & $\phi_{\mathrm{GW_U},3} (t-T_3/2)$  & $\phi_{\mathrm{M_1},3} (t)$ & $\phi_{\mathrm{M_2},3} (t)$ & $\phi_{\mathrm{GW_L},3} (t+T_3/2)$  & $\phi_{\mathrm{BS_L},3} (t+T_3)$ \\ \hline
    $\phi_4(t)$ & $\phi_{\mathrm{BS_U},4} (t-T_4)$ & $\phi_{\mathrm{GW_U},4} (t-T_4/2)$  & $\phi_{\mathrm{M_1},4} (t)$ & $\phi_{\mathrm{M_2},4} (t)$ & $\phi_{\mathrm{GW_L},4} (t+T_4/2)$  & $\phi_{\mathrm{BS_L},4} (t+T_4)$ \\  \hline
  \end{tabular}
  }
  \label{tab:1}
\end{table}

\noindent
Here, we assume that all terms carry a positive sign for simplicity. This is justified because in both paths the beams experience a reflection on the same side of the beam splitter, resulting in a common phase shift that cancels out.

Finally, we define the interference signal of neutrons of each speed as $s_\mathrm{i}=\phi_\mathrm{i}(t)/k_\mathrm{i}$. This interference signal indicates the sum of displacement noise or signal that neutrons of each speed experience, and thus it has a dimension of displacement. We also define the contribution of the interference signal caused by each component (mirror, BS, and GW), $s_\mathrm{l,i}(t)$, as $s_\mathrm{i}(t)= \sum s_\mathrm{l,i}(t)$. For example, the contribution from $\mathrm{BS_U}$ is $s_\mathrm{BS_U,i}(t)=\phi_\mathrm{BS_U,i}(t-T_{\mathrm{i}})/k_\mathrm{i}$.

Considering the situation without gravity, neutrons propagate straight and hit at the same point on each optical component. Since neutrons that strike the mirror simultaneously are subjected to identical mirror displacement noise, this noise can be canceled by combining pairs of signals. For example, the following combinations yield signals that are free from the mirror displacements:

\begin{align}
   s_{14} (t) = s_1 (t) - s_4 (t)  \label{eq:1} , & \\
   s_{23} (t) = s_2 (t) - s_3 (t) \label{eq:2} . & 
\end{align}

Now, we consider the interference signal in the frequency domain instead of the time domain to explain the cancellation of the displacement noises from $\mathrm{BS_U}$ and $\mathrm{BS_L}$. The Fourier spectrum of the interference signal is defined as $S_{\mathrm{x}}(\Omega) = \int_{-\infty}^{\infty} s_{\mathrm{x}}(t) e^{-j \Omega t} dt$, where $\Omega$ is the angular frequency, $j$ is an imaginary unit and x is arbitrary, representing i or a pair of i and l. Accordingly, the time-domain interference signal can be expressed as $s_{\mathrm{x}}(t) = \int_{-\infty}^{\infty} S_{\mathrm{x}}(\Omega) e^{j \Omega t} \, d\Omega$. Here, we impose a condition on the neutron speeds, which is given by the following equation (see \cite{DFNI_Iwaguchi_2} for more details; an intuitive explanation is provided later in this section).

\begin{equation}
    \frac{1}{v_1} + \frac{1}{v_4} = \frac{1}{v_2} + \frac{1}{v_3}.
    \label{eq:3}
\end{equation}

\noindent
This is based on selecting a speed such that the average of phase between $v_1$ and $v_4$ is the same as that between $v_2$ and $v_3$. This condition makes it possible to cancel the displacement noises of $\mathrm{BS_U}$ and $\mathrm{BS_L}$ with the DFI signal combination, which is written as

\begin{equation}
   S_\mathrm{DFI} (\Omega) = c_{14} (\Omega) S_{14} (\Omega) -  c_{23} (\Omega) S_{23} (\Omega),
    \label{eq:4}
\end{equation}

\noindent
with the coefficients

\begin{equation}
    c_{14} (\Omega) = \sin \Omega  \left(\frac{T_2 - T_3}{2} \right),   c_{23} (\Omega) = \sin \Omega \left( \frac{T_1 - T_4}{2} \right).
    \label{eq:5}    
\end{equation}

\begin{figure}[H]
   \centering
   \includegraphics[clip,width=0.45\textwidth]{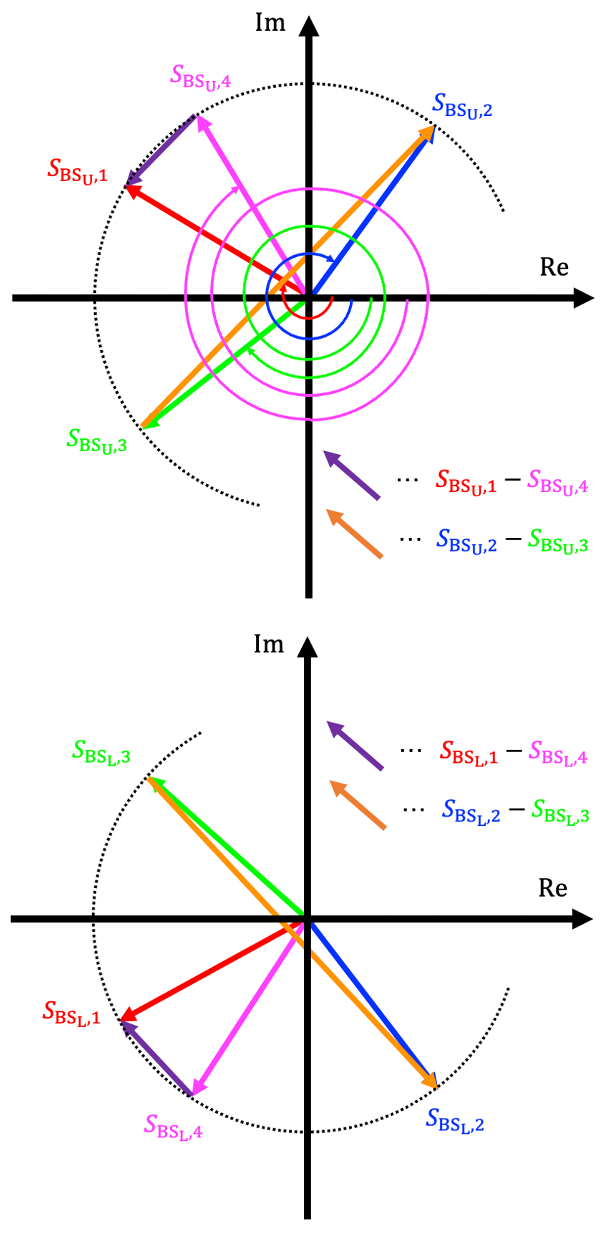}
   \caption{Phasor diagram of BS displacement noises. Solid arrows originating from the origin represent the displacement noises of $\mathrm{BS_U}$ (upper panel) and $\mathrm{BS_L}$ (lower panel). Arrow lengths correspond to the normalized amplitude of the displacement noise. Dashed circle shows the reference of normalized amplitude. Tangerine and purple arrows indicate the residual BS noises after all mirror noises are canceled in Eqs. \eqref{eq:1}-\eqref{eq:2}. In the upper panel, the length of the rotating arrows corresponds to the phase shift of BS displacement noises.}
   \label{fig:2a}
\end{figure}

\begin{figure}[H]
   \centering
   \includegraphics[clip,width=0.45\textwidth]{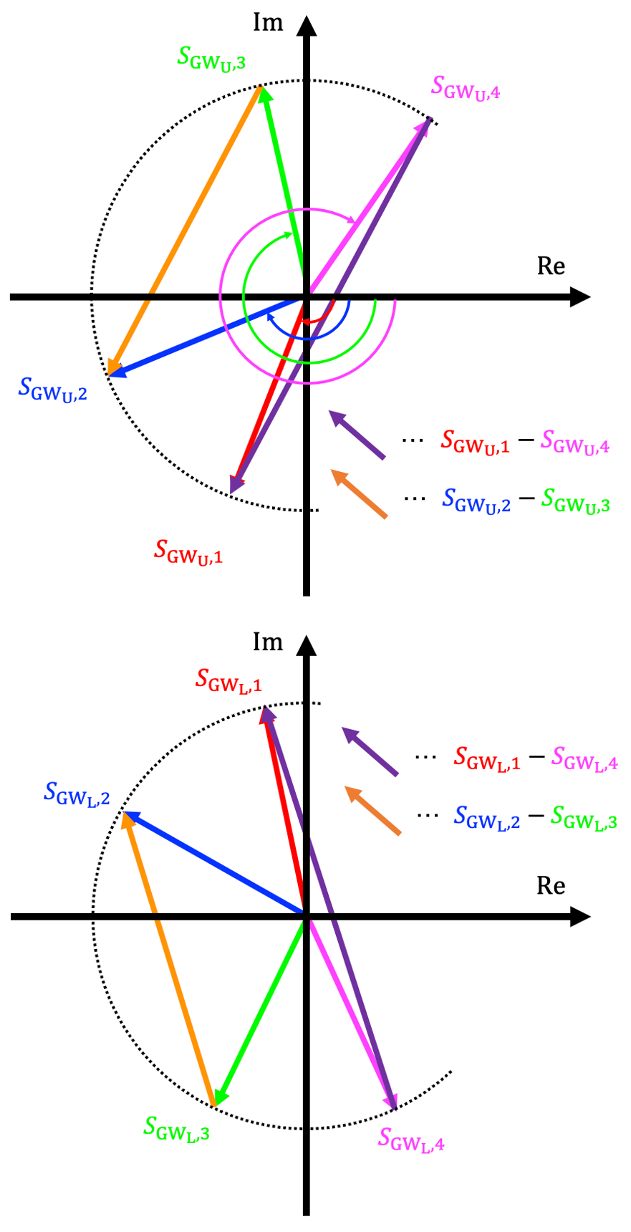}
   \caption{Phasor diagram of GW signals. Solid arrows originating from the origin represent the $\mathrm{GW_U}$ (upper panel) and $\mathrm{GW_L}$ (lower panel) signals. Arrow lengths correspond to the normalized amplitude of the GW signal. Dashed circle shows the reference of normalized amplitude. Tangerine and purple arrows indicate the residual GW signals after all mirror noises are canceled in Eqs. \eqref{eq:1}-\eqref{eq:2}. In the upper panel, the length of rotating arrows corresponds to the phase shift of GW signals.}
   \label{fig:2b}
\end{figure}

We use a phasor diagram to intuitively explain the cancellation of BS noises and the preservation of GW signals, as shown in Figs. \ref{fig:2a}-\ref{fig:2b}. In these figures, the Fourier component of the interference signal at $\Omega$, $S_{\mathrm{x}}(\Omega) e^{j \Omega t}$, where x indicates the source of the displacement or GW signal, is represented as arrows originating from the origin. For example, $S_\mathrm{BS_U,1}$ suggests the interference signal of neutrons with speed $v_1$, affected by the displacement noise of $\mathrm{BS_U}$. At the same time, $S_\mathrm{GW_L,2}$ indicates the interference signal of neutrons with speed $v_2$, affected by the GW signal in the lower part of the interferometer in Fig. \ref{fig:1}. Note that although the amplitude and initial phase of these noises or signals are arbitrary, they are assumed to be the same in these figures for simplicity. It is essential to recognize that the phase (indicated as the rotating arrows in the figures) is accumulated more for the BS noises than the GW signals, because of the difference in path length from the mirror. The tangerine and purple arrows correspond to the residuals of BS noises and GW signals after canceling the mirror noises by Eqs. \eqref{eq:1}-\eqref{eq:2}, which are free from any mirror displacement. The parallelism of these arrows results from the condition on the neutron speeds shown in Eq. \eqref{eq:3}.

The key to canceling the $\mathrm{BS_U}$ and $\mathrm{BS_L}$ noise components in Fig. \ref{fig:2a} — after the mirror displacement noise has been removed — lies in the fact that the tangerine and purple arrows are parallel. The $\mathrm{BS_U}$ and $\mathrm{BS_L}$ noise can be eliminated by subtracting these two vectors, weighted by coefficients $c_{14} (\Omega)$ and $c_{23} (\Omega)$ that account for their amplitude differences. In contrast to BS noises, the residual GW signals are preserved in Eq. \eqref{eq:4} because the length ratio of the arrows in the GW signals in Fig. \ref{fig:2b} is different from that in BS noises. In Fig. \ref{fig:2a}, the tangerine arrow is more than twice as long as the purple one and points in the opposite direction. In contrast, in Fig. \ref{fig:2b}, the tangerine arrow is slightly shorter than the purple one and points in the same direction. This difference illustrates that the GW signals are not canceled by the combination that eliminates the BS noises.

Here it should be emphasized that for the DFI combination, sufficient angular differences between the arrows in Figs. \ref{fig:2a}-\ref{fig:2b} are crucial, because they create differences in the length ratio between BSs noises and GW signals. This angular difference, which corresponds to the difference of phase shifts in terms of noise and signal, is the essence of DFI.

\section{Comparison between ideal and actual experiment}
\label{section:3}

\begin{figure}[H]
   \centering
   \includegraphics[clip,width=0.45\textwidth]{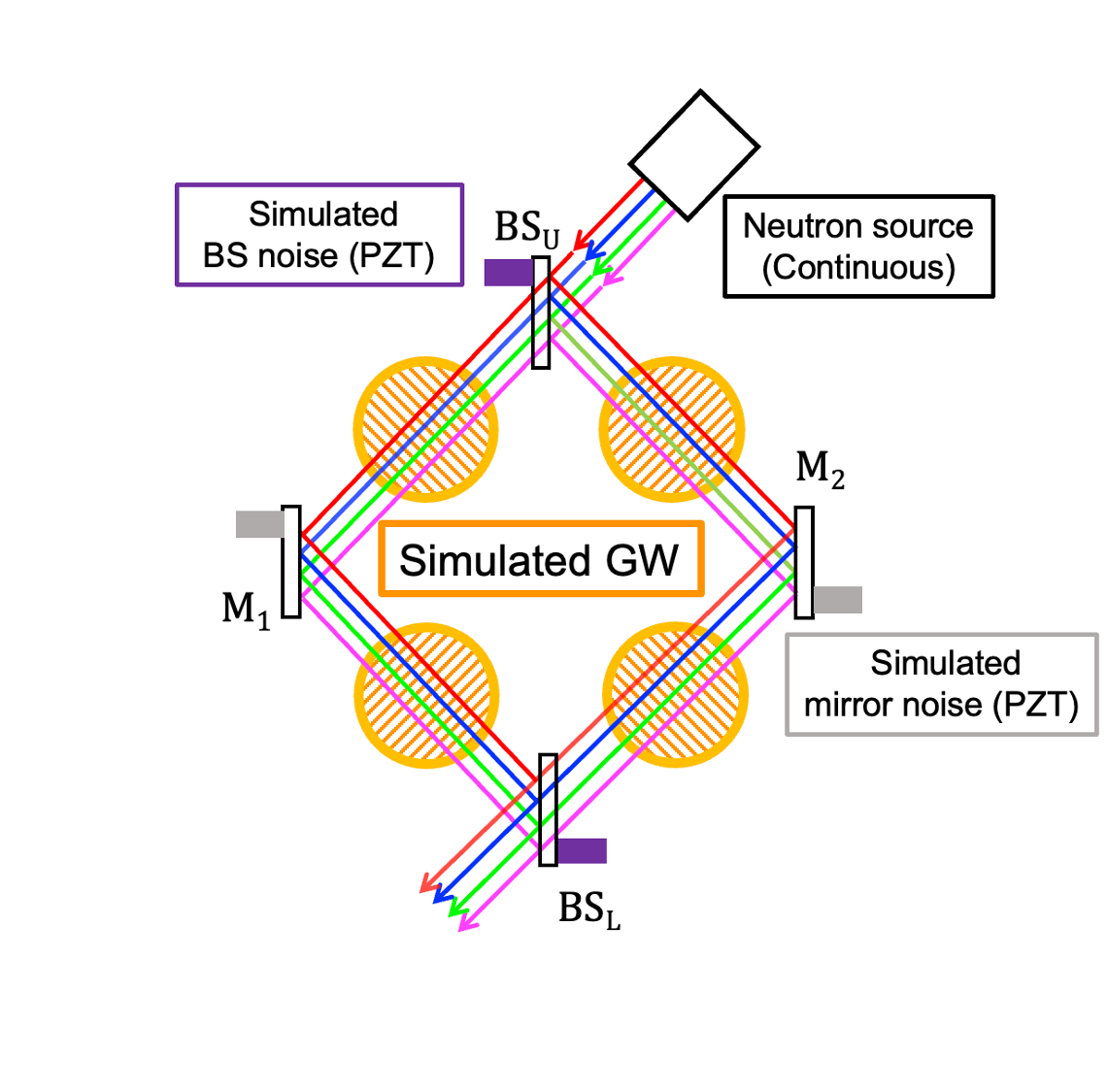}
   \caption{Straightforward experimental setup for the ideal proof-of-principle experiment on the four-speed unidirectional neutron DFI. The red, blue, green, and magenta arrows indicate the trajectories of neutrons with the speeds of $v_\mathrm{1}$, $v_\mathrm{2}$, $v_\mathrm{3}$, and $v_\mathrm{4}$, respectively. The purple and grey rectangles represent actuation systems for BS noises and mirror noises. The mesh circles colored yellow show actuation systems for GW signals.}
   \label{fig:3}
\end{figure}

To illustrate the conceptual foundation of our work, Figure \ref{fig:3} depicts an idealized setup for a proof-of-principle experiment involving a four-speed, unidirectional neutron DFI. This setup comprises a square Mach-Zehnder interferometer illuminated by a continuous, high-intensity neutron beam. The high neutron flux suppresses shot noise, thereby improving signal clarity. Mirrors and BSs in this configuration are equipped with actuators (e.g., piezoelectric transducers) to simulate displacement noise. Additional phase modulation elements between optical components are used to simulate GW signals.

In this ideal configuration, both simulated displacement noise and GW signals are introduced, enabling validation of the DFI principle: appropriate combinations of interferometric outputs can cancel displacement noise while preserving GW signals. The neutron beams are assumed to strike mirrors simultaneously, which allows the mirror noise cancellation in  Eqs. \eqref{eq:1}-\eqref{eq:2}, as discussed in Section \ref{section:2}, and the interferometer's arm lengths and modulation frequencies are selected to ensure sufficient phase differences, typically on the order of $\pi$ between neutrons of different speeds. As established in Section \ref{section:2}, this phase difference corresponds to angular separation in the phasor diagram, which is essential for differentiating noise from GW signals. Therefore, the experimental setup must be designed with appropriate modulation depth for the simulated signal and noise and a suitable path length of the interferometer.

However, our experimental setup deviates from this ideal due to several practical constraints. As shown in Fig. \ref{fig:4}, the experiment was carried out at the Materials and Life Science Experimental Facility (MLF) at J-PARC \cite{J-PARC}, using equipment based on an earlier neutron interferometer demonstration \cite{Fujiie_PRL} that successfully employed beam-splitting etalons (BSEs).

\begin{table}[H]
   \centering
   \begin{minipage}{0.22\textwidth}
       \centering
       \includegraphics[clip,width=\textwidth]{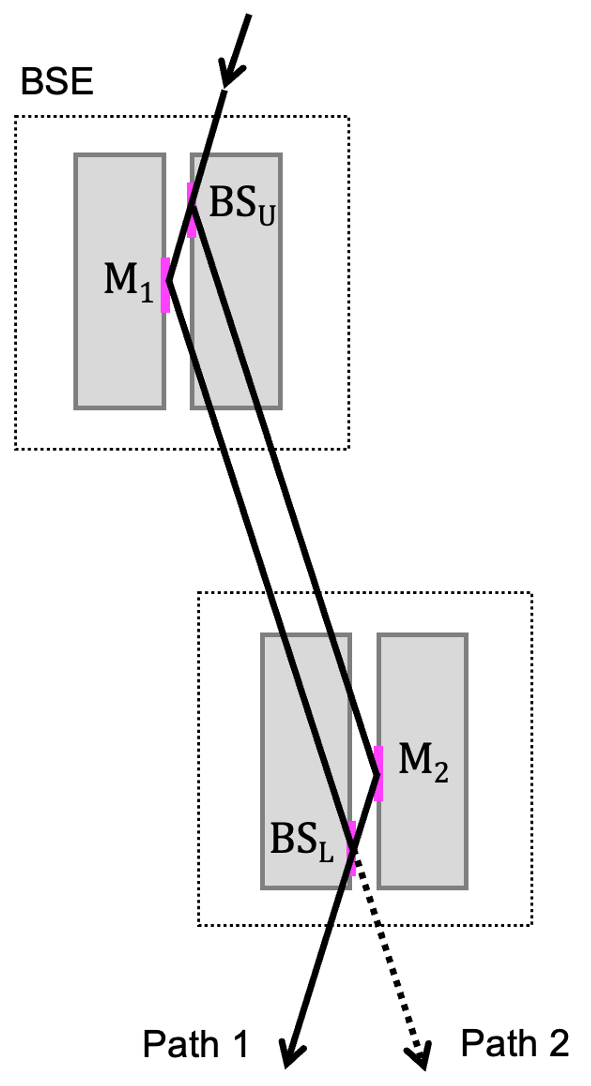}
   \end{minipage}
   \begin{minipage}{0.23\textwidth}
       \centering
       \includegraphics[clip,width=\textwidth]{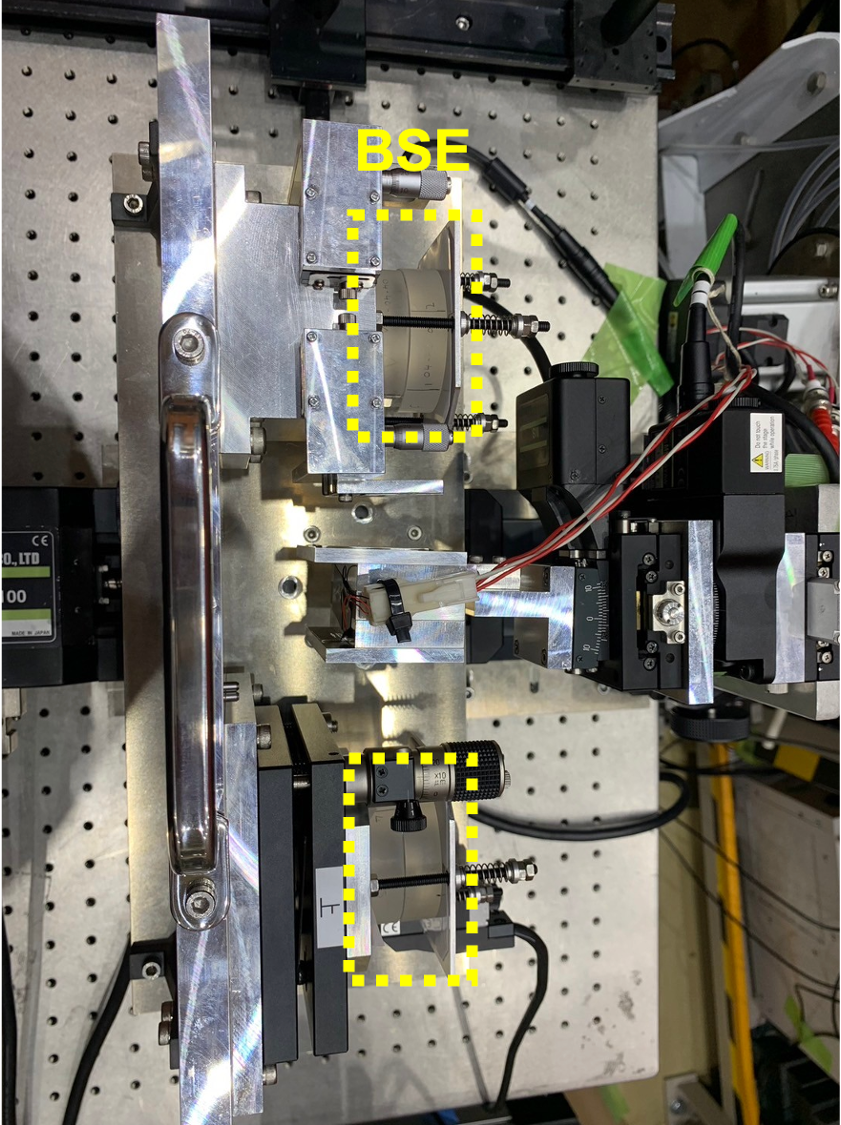}
   \end{minipage}
   \captionof{figure}{Actual setup for the neutron DFI used at J-PARC (left panel), and a photograph of the setup with the inserted sample displayed (right panel). In the left panel, neutron mirrors and BSs correspond to magenta lines and the paths of interfering neutrons are illustrated by the solid (Path 1) and dashed (Path 2) black arrows. In the photograph of the right panel, two beam-splitting etalons (BSEs) are highlighted by yellow squares, and the neutron beam enters there from the top side of the picture.}
   \label{fig:4}
\end{table}

The interferometer adopts a parallelogram geometry with an acute angle of approximately $1\tcdegree$, instead of a square. Instead of separate mirrors and BSs with independent actuators, two BSEs are used \cite{BSE_Kitaguchi, BSE_Seki}. These BSEs consist of two super-polished $\mathrm{SiO_2}$ plates facing each other with fixed spacers. Consequently, the mirrors and BSs are physically integrated and cannot be actuated independently, unlike in the ideal setup.

Another significant limitation lies in the available path length. Achieving sufficient phase modulation requires a long path relative to the modulation frequency; for example, a path length of approximately 1.25 m is required to obtain half-wave phase modulation at a modulation frequency of 200 Hz for neutrons traveling at 500 m/s. However, the practical path length is limited to approximately 0.2 meters, constrained by alignment tolerances of the BSEs. This short path length limits the achievable phase difference between neutrons of different speeds to below $\pi$.

The neutron source at J-PARC’s BL05 beamline is a pulsed spallation source, particularly the low-divergence branch of the  BL05 (NOP) \cite{J-PARC}. For the neutron DFI to function, two conditions must be met: (1) clear separation of neutrons of different speeds, and (2) simultaneous arrival of neutrons at the mirrors. A continuous source would naturally fulfil the second condition and could separate neutron speeds using a spectral dispersion prism. In contrast, the pulsed source allows speed selection via time-of-flight (TOF) measurements but makes it challenging to ensure simultaneous mirror arrival.

The neutron flux is also considerably lower than in the ideal scenario. At a wavelength of 1 nm, the BL05 beamline provides a flux of approximately $10^2$ neutrons per second—far below the $10^{13}$ neutrons per second expected from a continuous source with shorter wavelengths (e.g., 0.13 nm) \cite{ESS}.

These differences between the ideal and actual setups are summarized in Table \ref{tab:2} below:

\begin{table}[h]
  \centering
  \caption{Differences between the ideal and actual setups.}
  \scalebox{0.625}{
  {\renewcommand\arraystretch{2.0}
  \begin{tabular}{l||c|c} \hline
    Category & Ideal setup & Actual setup \\  \hline \hline
    \ctext{1} Geometry & Square interferometer & Acute-angled parallelogram \\ \hline
    \ctext{2} Components & Two independent mirrors and BSs & Integrated BSEs (non-actuable) \\ \hline
    \ctext{3} Neutron source & Continuous beam & Pulsed beam \\ \hline
    \ctext{4} Phase difference & $ \sim \pi$ & $ \ll \pi$ \\ \hline
    \ctext{5} Neutron beam flux &  $ \sim 10^{13} \ / \mathrm{s} $ (for $\lambda = 0.13$ $\mathrm{nm}$)  & $ \sim 10^{2} \ / \mathrm{s} $ (for $\lambda = 1$ $\mathrm{nm}$)\\ \hline
  \end{tabular}}}
  \label{tab:2}
\end{table}

In the next section, we describe the methods employed to compensate for these differences, enabling the realization of a functional proof-of-principle neutron DFI experiment using the constrained apparatus.

\section{Details of actual setup, Analysis method and numerical simulation result}
\label{section:4}

This section outlines the experimental design and analysis methodology used to replicate, as closely as possible, the essential behaviors of the ideal DFI configuration within the constraints of our available setup. This proof-of-principle experiment aims to demonstrate that BS displacement noise can be effectively canceled using DFI signal processing, while simulated GW signals are preserved.

\subsection{Experimental Apparatus}
\label{subsection:0}

Fig. \ref{fig:5} shows the experimental layout at J-PARC. A pair of aluminum (Al) phase plates—used to induce neutron phase shifts—is inserted at a $45\tcdegree$ angle into both neutron paths. The pair consists of a modulator and a compensator. The compensator offsets the beam path displacement introduced by the modulator. Each Al plate features a zigzag shape to reduce spatial requirements while maintaining the $45\tcdegree$ insertion angle. The thickness of the plates along the beam path is 6 mm.

\begin{table}[H]
   \centering
   \begin{minipage}{0.217\textwidth}
       \centering
       \includegraphics[clip,width=\textwidth]{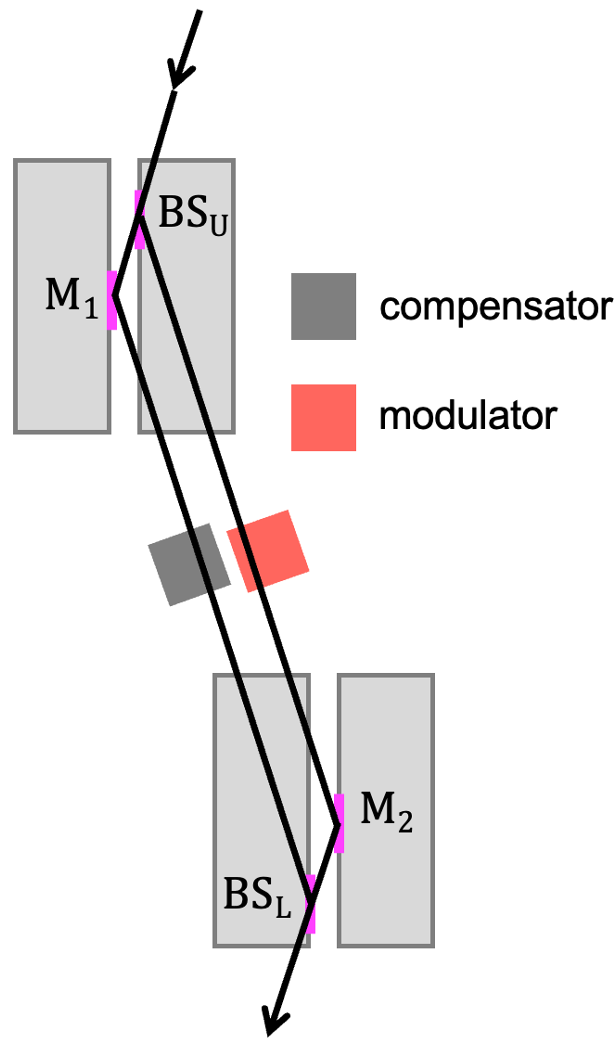}
   \end{minipage}
   \begin{minipage}{0.23\textwidth}
       \centering
       \includegraphics[clip,width=\textwidth]{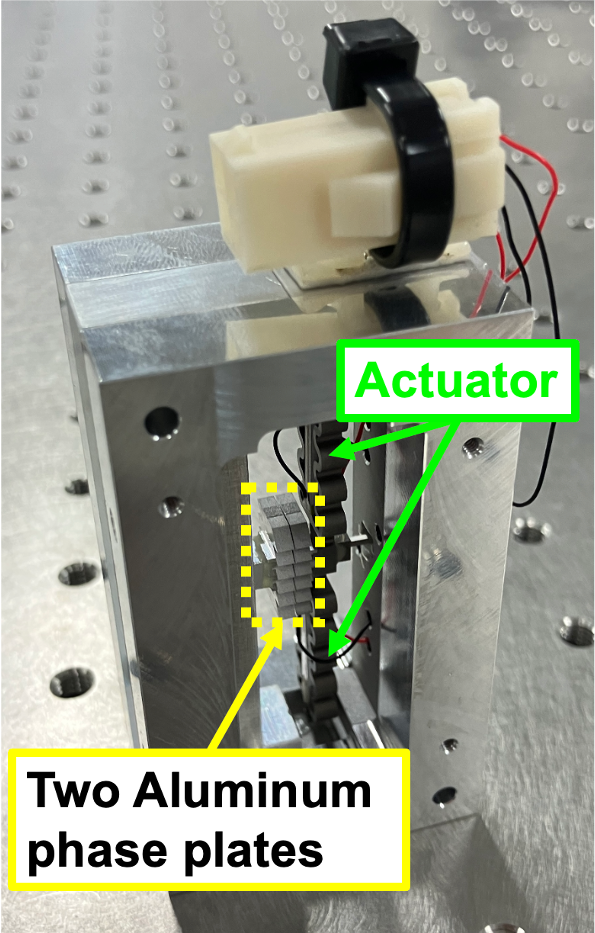}
   \end{minipage}
   \captionof{figure}{Actual setup for the proof-of-principle experiment (left panel) and photograph of an inserted sample (right panel). In the picture in the right panel, a pair of Al phase plates of modulator and compensator is highlighted by the yellow square and the actuators (piezoelectric transducers) are indicated by the green arrows.}
   \label{fig:5}
\end{table}

The modulator plate is actuated using a piezoelectric transducer, which varies the plate's angle and thus the effective flight path length through Al relative to air. Because Al and air have different neutron refractive indices, this results in phase modulation of the neutron interference signal. The modulation is synchronized with the neutron source trigger, enabling controlled phase shifts in the neutron pulses.

Due to the limited space between the BSEs ($\sim$50 mm), only one actuator can be placed along the beam path. This constraint necessitated the development of a novel analysis method, detailed below, to simulate the independent effects of mirror noise, BS noise, and GW signals—each of which ideally would be applied at different physical locations.

\subsection{Simulated signals of GWs and displacement noises}
\label{subsection:1}

To simulate the effects of GW signals and displacement noise (excluding mirror noise, which is theoretically cancelable via signal combinations as shown in Section \ref{section:2}), we apply five modulation patterns:

\begin{align}
   & 1. \mathrm{BS_U} \ \mathrm{noise} \notag \\
   & 2. \mathrm{BS_L} \ \mathrm{noise} \notag \\
   & 3. \mathrm{GW_U} \ \mathrm{signal} \notag \\
   & 4. \mathrm{GW_L} \ \mathrm{signal} \notag \\
   & 5. \mathrm{Mirror} \ \mathrm{noise} \ (\text{used as a timing reference}) \notag
\end{align}

In the noise simulation, the cancellation target is only BS noise because the cancellation of mirror noises is theoretically evident as shown in Eqs. \eqref{eq:1} and \eqref{eq:2}. This adjustment corresponds to difference \ctext{1} (Geometry) listed in Table \ref{tab:2}. The mirror noise is used as a timing reference for BS noise cancellation.

\begin{figure}[H]
   \centering
   \includegraphics[clip,width=0.45\textwidth]{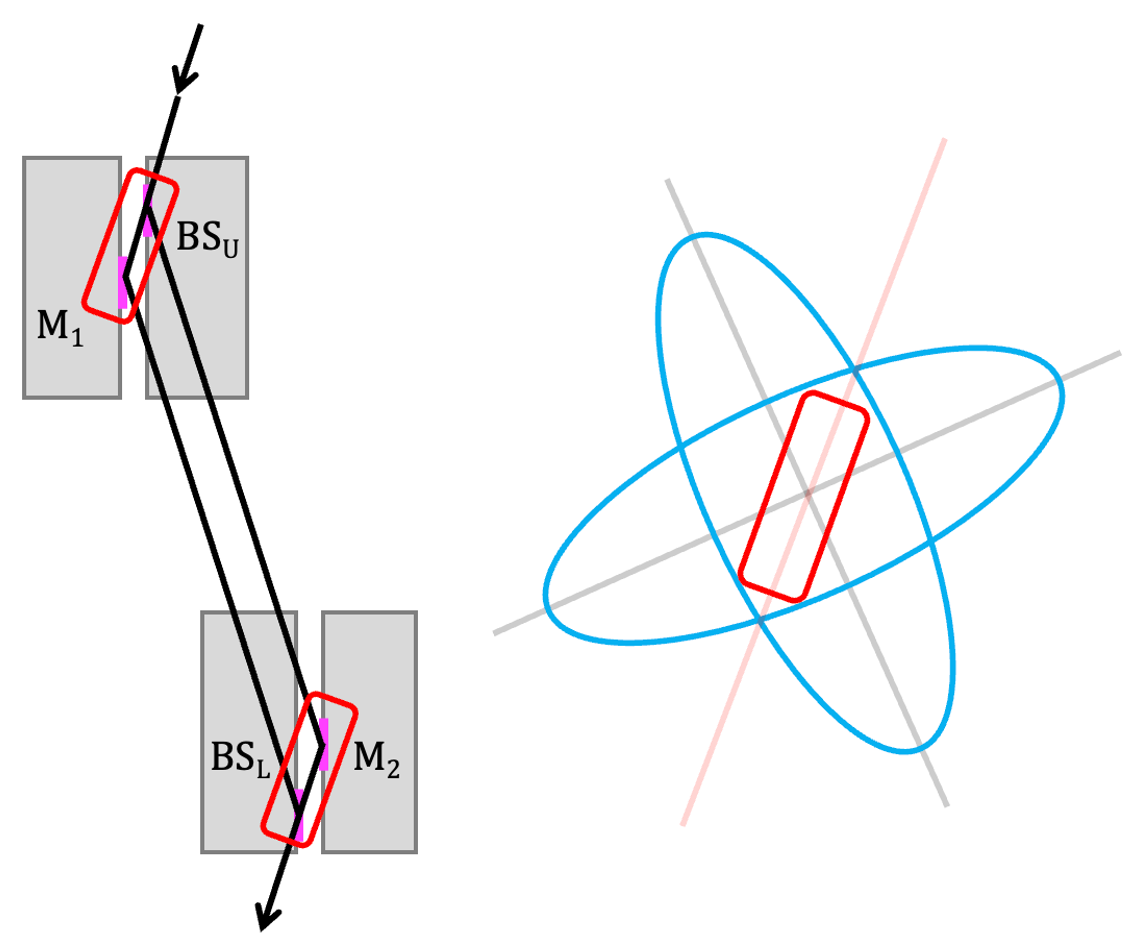}
   \caption{GW polarization with no effect on the shorter paths. The region indicated by the red rectangle is where the GW effect can be ignored when the GW propagating in the z direction has the polarization represented by the blue elliptical shape.}
   \label{fig:6}
\end{figure}

In the GW simulation, the shorter arm of the interferometer lies within the beam-splitting etalon (BSE), as shown in Fig. \ref{fig:6}, making it challenging to apply simulated GW signals directly to this path. To accommodate GW polarization without influencing the shorter arm, we simulate a cross-polarization mode oriented at a 45-degree angle relative to the shorter path, as illustrated in Fig. \ref{fig:6}. Consequently, in the demonstration setup, simulated GW signals are applied only to the longer arms of the interferometer. This adjustment corresponds to hardware limitation \ctext{2} (components) listed in Table \ref{tab:2}.

A phase plate is inserted at the mid-point between the two BSEs since piezoelectric transducers cannot be directly mounted  on the BSEs. As a result, in this setup, the simulated BS noise includes a phase shift offset that arises from the spatial separation between the actual BS and the location of the phase plate. The simulated GW signals are considered as the integrated sum of the phase shift during neutron propagation between the two BSEs. As summarised in Table II, these adaptations also stem from hardware limitation \ctext{2} (components).

\subsection{Demonstration Setup and Analysis for the DFI} 
\label{subsection:2}

In this subsection, we describe the analysis methods developed for the actual setup and present simulation results, as a precursor to the experimental data analysis in Section \ref{section:5}.

Four neutron speeds were selected at BL05: $v_1 = 467.3$ m/s, $v_2 = 435.0$ m/s, $v_3 = 381.7$ m/s, $v_4 = 360.0$ m/s, all of which satisfy the DFI velocity condition defined in Eq. \ref{eq:3}. In practice, these neutron speeds are identified using time-of-flight (TOF) measurements, which typically carry an uncertainty of ±50 $\mu$s. The TOF uncertainty originates from the finite moderator size, causing a time spread in neutron velocity. This mainly reduces interference visibility, but its effect is negligible because it is mitigated by statistical accumulation. For the simulated data, we consider five types of signals:  $\mathrm{BS_U}$, $\mathrm{GW_U}$, $\mathrm{M}$ (mirror noise), $\mathrm{GW_L}$, and $\mathrm{BS_L}$. Among these, the mirror noise signal is assigned a frequency of $f_{\mathrm{M}} = 200$ Hz, which serves as the time reference for canceling the displacement noise originating from the BSs.

\begin{figure}[H]
   \centering
   \includegraphics[clip,width=0.49\textwidth]{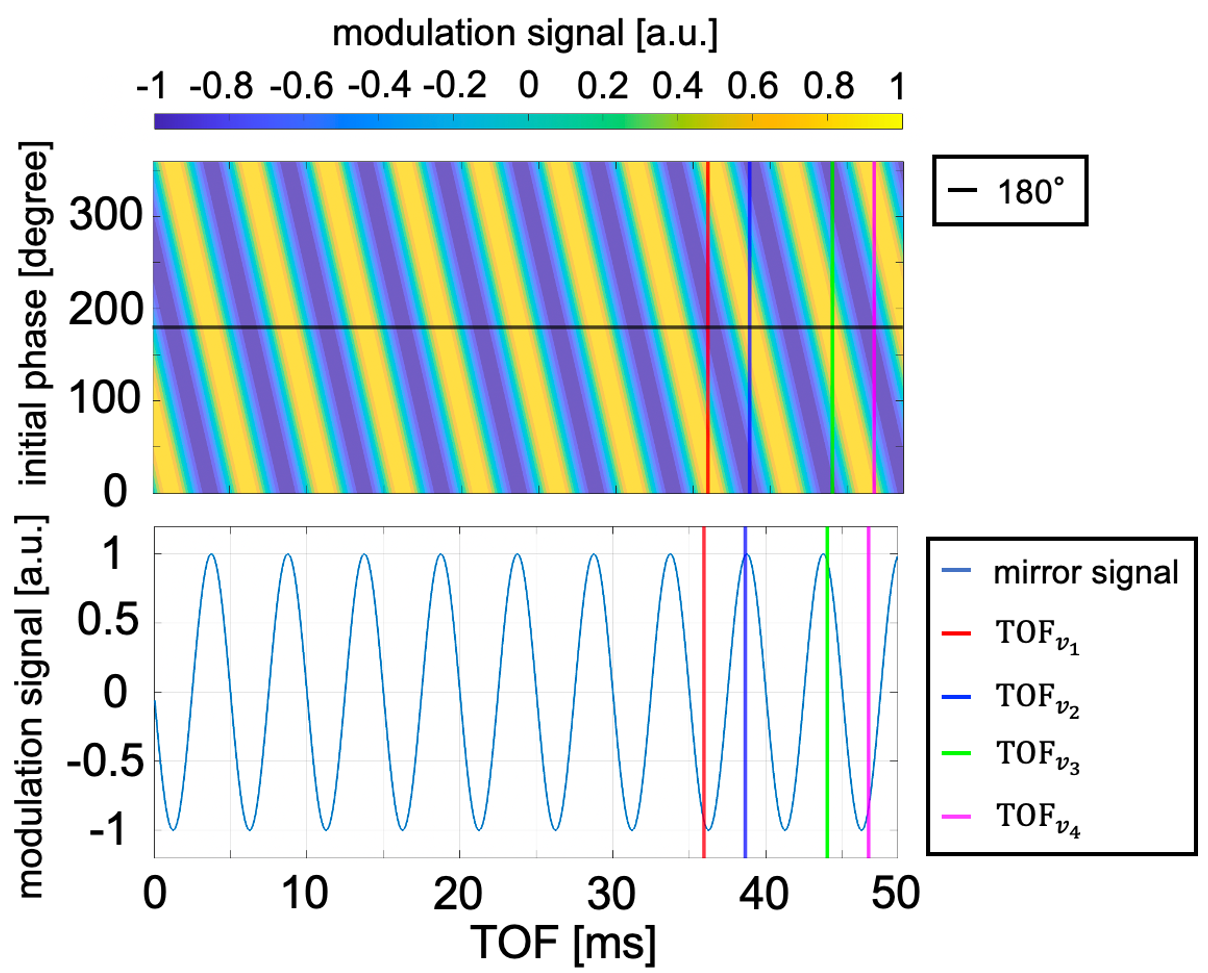}
   \caption{Simulated modulation signals for the mirror noise at 200 Hz. The upper panel shows the modulation signals as a function of the TOF of neutrons and the initial phase of the modulation signal. The lower panel shows the modulation signals as a function of the TOF of neutrons at an initial phase of 180 degrees. The colored lines indicate the TOFs corresponding to the four neutron speeds used in the experiment. Both panels share a common horizontal axis.}
   \label{fig:7}
\end{figure}

In conventional laser interferometers, mirror modulations manifest as variations in the interference fringe pattern, which depends on the difference in the optical path length between the two arms.  In contrast, in our neutron interferometer, the interference fringes are represented as a function of the TOF, as shown in Fig. \ref{fig:7}. This TOF-based representation is particularly advantageous in our case, given the low neutron beam flux (see Table  \ref{tab:2}), which necessitates long accumulation times to achieve adequate fringe contrast. Under these experimental conditions, TOF provides a more practical and informative framework for analyzing the interference signals.

The interference fringes of the neutron interferometer, $I(\lambda)$, are defined by the following formulas:

\begin{equation}
    I(\lambda) = \frac{I_\mathrm{Path 1}/I_{\mathrm{Path 1}, \mathrm{SinglePass}} - I_\mathrm{Path 2}/I_{\mathrm{Path 2}, \mathrm{SinglePass}}}{I_\mathrm{Path 1}/I_{\mathrm{Path 1}, \mathrm{SinglePass}} + I_\mathrm{Path 2}/I_{\mathrm{Path 2}, \mathrm{SinglePass}}},
    \label{eq:15}
\end{equation}

\noindent
where $I_\mathrm{path 1}$ and $I_\mathrm{path 2}$ are the neutron counts for the beams of path 1 and path 2 (shown in Fig. \ref{fig:4}), respectively, and $I_{\mathrm{path 1}, \mathrm{SinglePass}}$ and $I_{\mathrm{path 2}, \mathrm{SinglePass}}$ are the neutron counts for each path when the other path is blocked by a Cd slit \cite{Fujiie_PRL}. Thus, $I(\lambda)$ ranges at most from -1 to +1, depending on neutron's wavelength (and thus the TOF). It is important to note that the measured interference fringes, a function of neutron wavelength, depend on the optical path length offset within the interferometer. The modulation signals discussed later in this section are derived from these measured interference fringes.

In the upper panel of Fig. \ref{fig:7}, the horizontal axis represents the TOF of neutrons, corresponding to the neutron speed (and thus, wavelength). The vertical axis represents the initial phase of the simulated signal. To ensure consistent modulation effects during data accumulation. The modulation signal is synchronized with the trigger produced by the neutron pulse generator to ensure consistent modulation effects throughout the data accumulation process.

\begin{figure}[H]
   \centering
   \includegraphics[clip,width=0.48\textwidth]{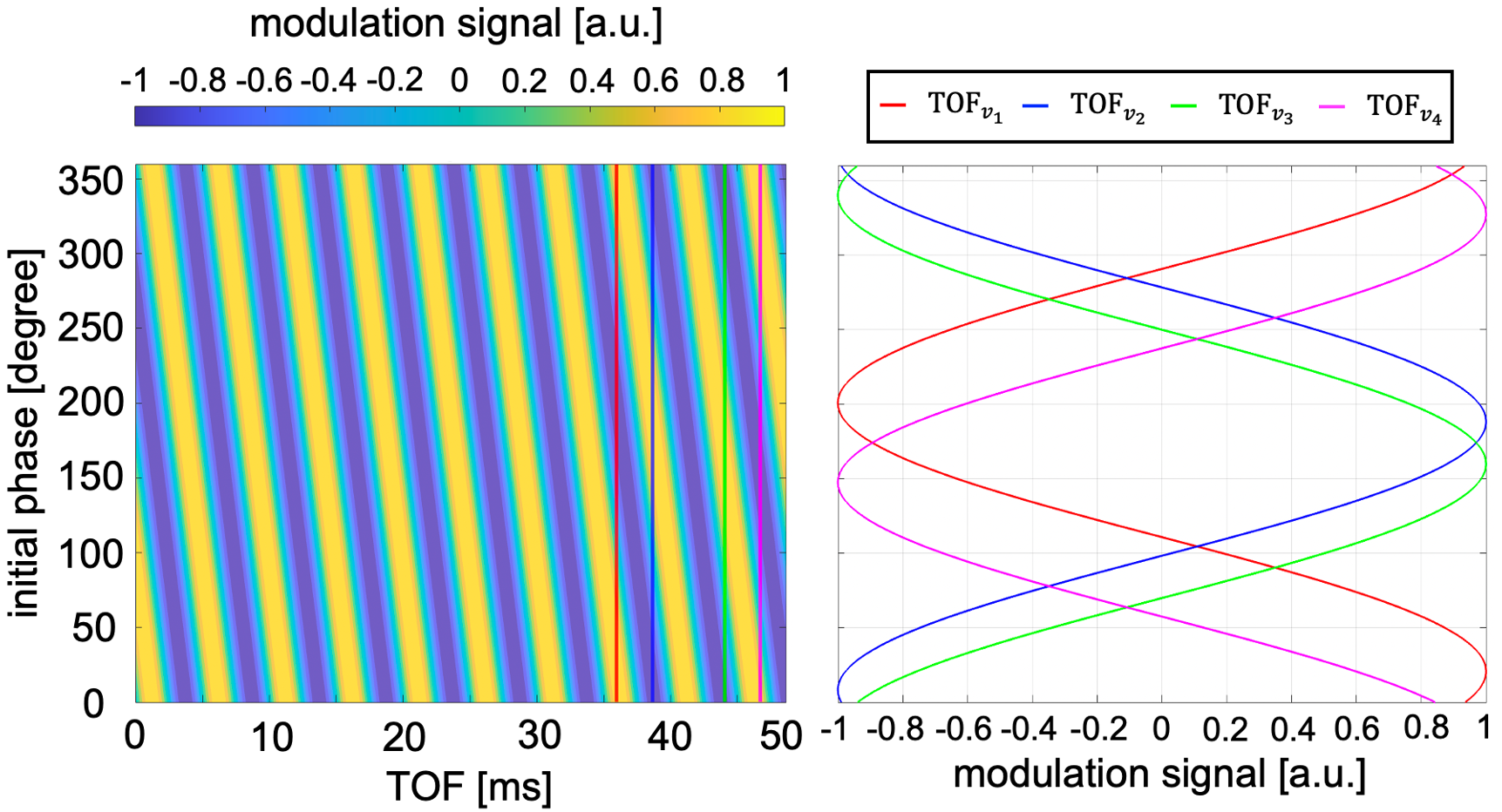}
   \caption{Simulated modulation signals for the mirror noise at 200 Hz for neutrons of different speeds. The left panel is the same as the upper panel in Fig. \ref{fig:8}. The right panel shows the modulation signals for neutrons of each speed as a function of the initial phase at the TOF corresponding to each neutron speed. Both panels share a common vertical axis.} 
   \label{fig:8}
\end{figure}

To observe how interference fringes vary with TOF, the initial phase of the modulation signal is systematically swept through a full 360 degrees during each measurement cycle — a procedure referred to as a phase scan. The lower panel of Fig. \ref{fig:7} displays modulation signals as a function of TOF at a fixed initial phase of 180 degrees, serving as a representative result from the phase scan. For subsequent numerical analysis, phase variation data are extracted for each neutron velocity, as illustrated in Fig. \ref{fig:8}.

The left panel of Fig. \ref{fig:8} replicates the upper panel of Fig. \ref{fig:7}, displaying the simulated modulation signals as a function of neutron TOF and the initial phase. In the right panel, the modulation signals are plotted for each neutron speed as a function of the initial phase, with the TOF fixed at the value corresponding to that specific speed.

\begin{figure}[H]
   \centering
   \includegraphics[clip,width=0.48\textwidth]{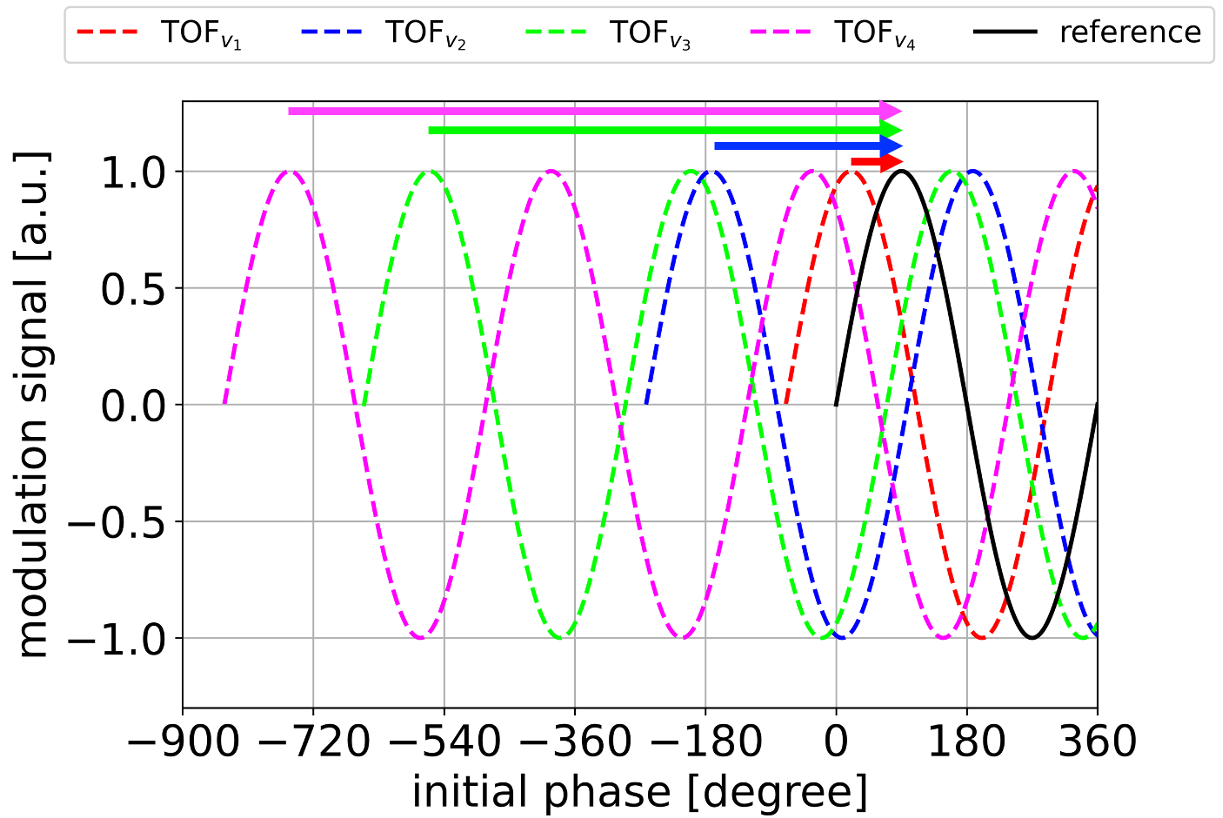}
   \caption{Simulated modulation signals as a function of the initial phase before (the colored dashed curves) and after (the black curves) the compensation. The size and direction of the colored arrows indicate the amount of correction.}
   \label{fig:9}
\end{figure}

In addition to the phase scan, a phase compensation is applied during the actual measurement as parts of data post-processing to simulate a condition where all neutrons arrive at the mirror with the same modulation phase. This compensation equalises the phase of the mirror modulation received by the four types of neutrons. The mirror modulation signal at an initial phase of zero is used as a reference for the compensation, ensuring that neutrons reach the mirror in synchrony with the modulation phase of the mirror. This phase compensation corresponds to difference \ctext{3} (Neutron source) listed in Table \ref{tab:2}. The modulation signals as a function of the initial phase, before and after the compensation, are shown in Fig. \ref{fig:9}.

In this figure, the black curve represents the reference. The colored curves show the modulation signals for neutrons of each speed (as in the right panel of Fig. \ref{fig:8}). The colored arrows indicate the amount of phase compensation applied to each neutron speed. The phase compensation derived from this simulation is applied to the experimental data to account for the variation in the timing of neutron arrivals times at the mirror.

A critical issue arises before applying this phase compensation to the phase variation data for both the BS noise and the GW signal simulation. Due to the limited space (approximately 50 mm) between the two BSEs, only one actuator can be inserted. Additionally, the modulation frequency for the actuators is limited to below 300 Hz to avoid heat generation during actuation. Even if multiple actuators could be placed along the path, the propagation time for each neutron between actuators (corresponding to the BSs and GWs) would still be much shorter than the modulation period. As a result, the phase differences between modulation signals at each actuator would be too small to achieve the DFI combination while preserving the GW signals successfully.

To address this issue, we use different modulation frequencies for each actuator — mirrors, BSs, and GWs. This adjustment is necessary because the difference in modulation frequencies corresponds to the phase shift caused by BS noises or GW signals, allowing for a clear separation between these signals despite the shared path. The frequency offset for BSs and GWs are set as $\Delta f_\mathrm{GW} = 50$ Hz and $\Delta f_\mathrm{BS} = 100$ Hz, with the modulation frequency for the mirror being 200 Hz ($f_\mathrm{M}$). Thus, BS and GW frequencies are as follows: 

\begin{align}
   & f_{\mathrm{BS_U}} = f_{\mathrm{M}} - \Delta f_\mathrm{BS} = 100 \mathrm{Hz}, \notag \\
   & f_{\mathrm{GW_U}} = f_{\mathrm{M}} - \Delta f_\mathrm{GW} = 150 \mathrm{Hz}, \notag \\
   & f_{\mathrm{GW_L}} = f_{\mathrm{M}} + \Delta f_\mathrm{GW} = 250 \mathrm{Hz}, \notag \\
   & f_{\mathrm{BS_L}} = f_{\mathrm{M}} + \Delta f_\mathrm{BS} = 300 \mathrm{Hz}. \notag
\end{align}

\noindent
These frequency differences result in the phase shift corresponding to the path length of 8.4 m.

In this experiment, each simulated signal for the displacement noise of the optics and the GW signals is applied individually rather than simultaneously as a combined signal. This approach is taken due to the upper limit of the piezoelectric voltages and the sufficiency of individual modulations.

All these measurements are then analyzed comprehensively in the post-processing stage. These methodologies account for the difference \ctext{4} (path length) in Table \ref{tab:2}. 

The modulation signals for each component are defined as follows:

\begin{align}
    \phi_\mathrm{BS_U} (t) &= \sin (  2 \pi f_{\mathrm{BS_U}} t ) \label{eq:6} , \\
    \phi_\mathrm{GW_U} (t) &= \sin (  2 \pi f_{\mathrm{GW_U}} t ) \label{eq:7} , \\
    \phi_\mathrm{M} (t)    &= \sin (  2 \pi f_{\mathrm{M}} t ) \label{eq:8} , \\
    \phi_\mathrm{GW_L} (t) &= \sin (  2 \pi f_{\mathrm{GW_L}} t ) \label{eq:9} , \\
    \phi_\mathrm{BS_L} (t) &= \sin (  2 \pi f_{\mathrm{BS_L}} t )  \label{eq:10} .    
\end{align}

\begin{figure}[H]
   \centering
   \includegraphics[clip,width=0.48\textwidth]{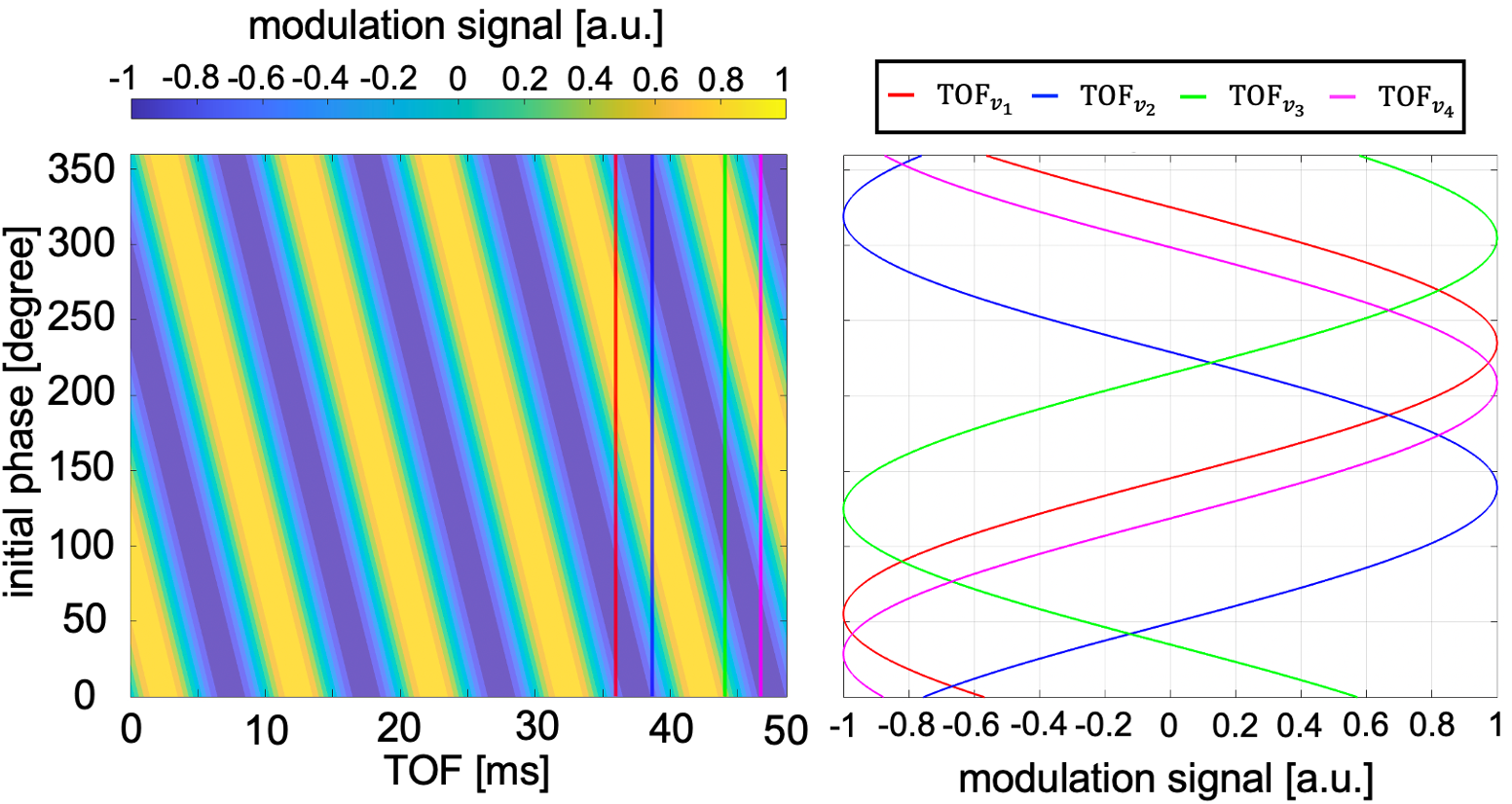}
   \caption{Simulated modulation signals for the $\mathrm{BS_U}$ noise at 100 Hz. The left panel shows the modulation signals as a function of the TOF of neutrons and the initial phase of the modulation signal. The right panel shows the modulation signals for neutrons of each speed as a function of the initial phase at the TOF corresponding to each neutron speed. Both panels share a common vertical axis.}
   \label{fig:10}
\end{figure}

\begin{figure}[H]
   \centering
   \includegraphics[clip,width=0.48\textwidth]{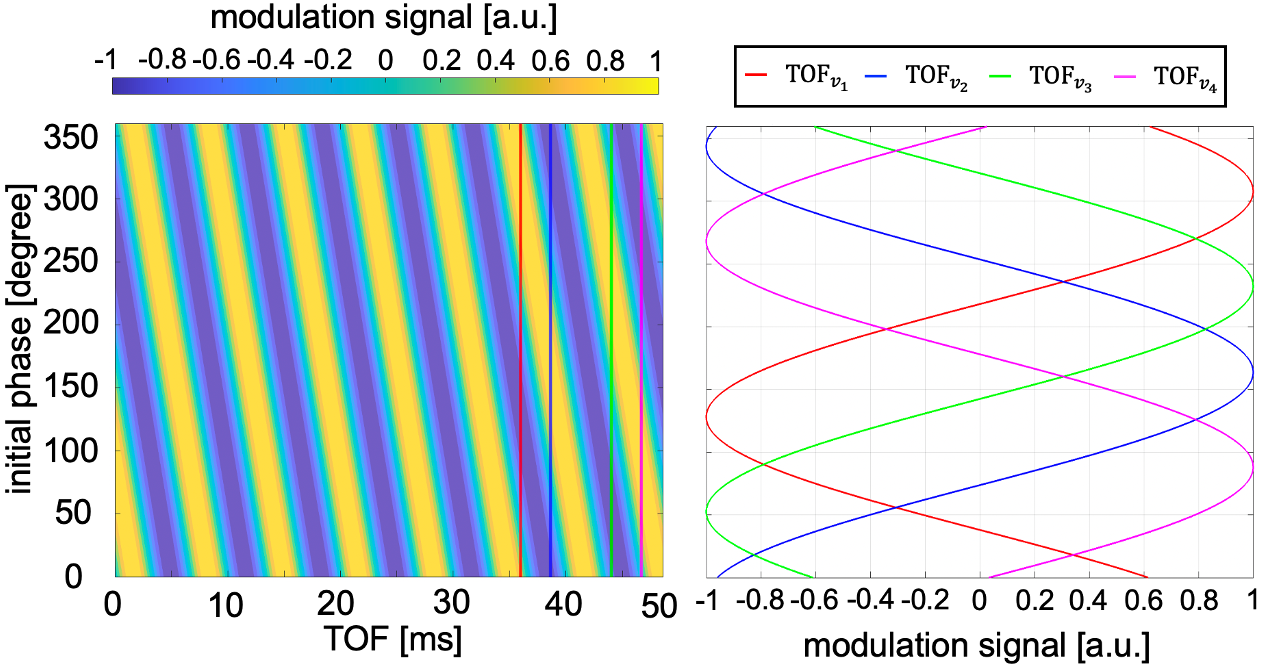}
   \caption{Simulated modulation signals for the $\mathrm{GW_U}$ signal at 150 Hz. All parameters and definitions in this figure are the same as those in Fig. \ref{fig:10}.}
   \label{fig:11}
\end{figure}

\begin{figure}[H]
   \centering
   \includegraphics[clip,width=0.48\textwidth]{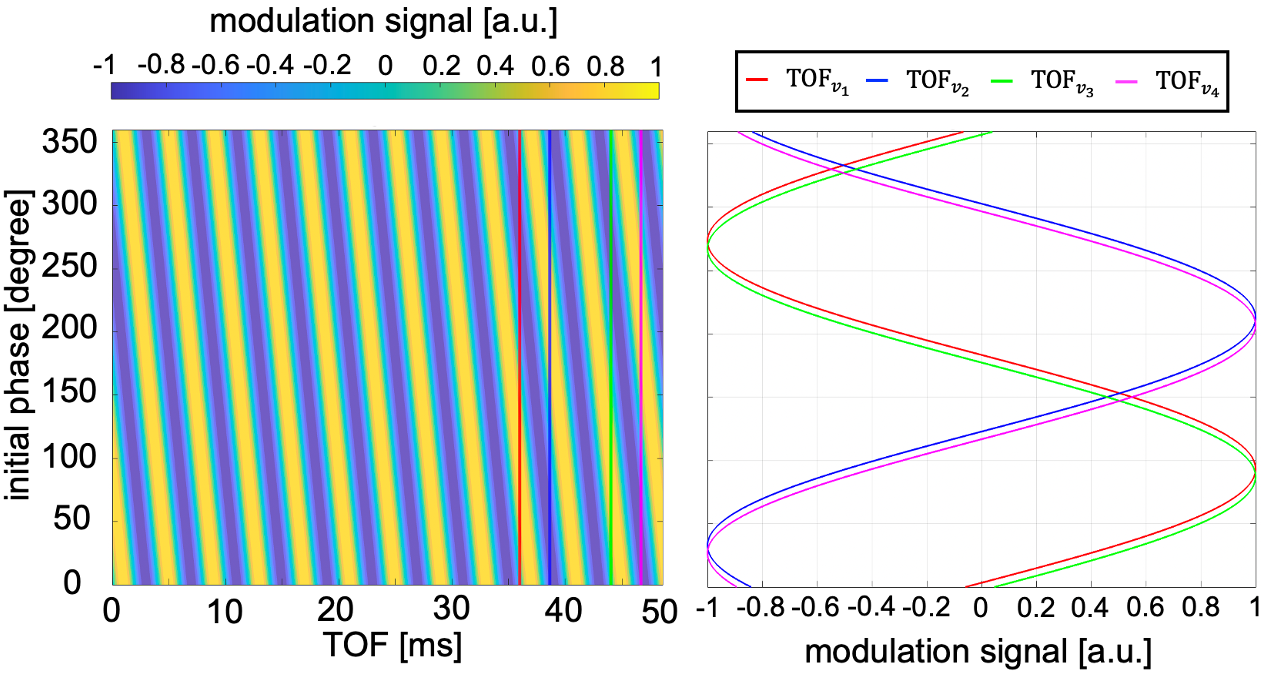}
   \caption{Simulated modulation signals for the $\mathrm{GW_L}$ signal at 250 Hz. All parameters and definitions in this figure are the same as those in Fig. \ref{fig:10}.}
   \label{fig:12}
\end{figure}

\begin{figure}[H]
   \centering
   \includegraphics[clip,width=0.48\textwidth]{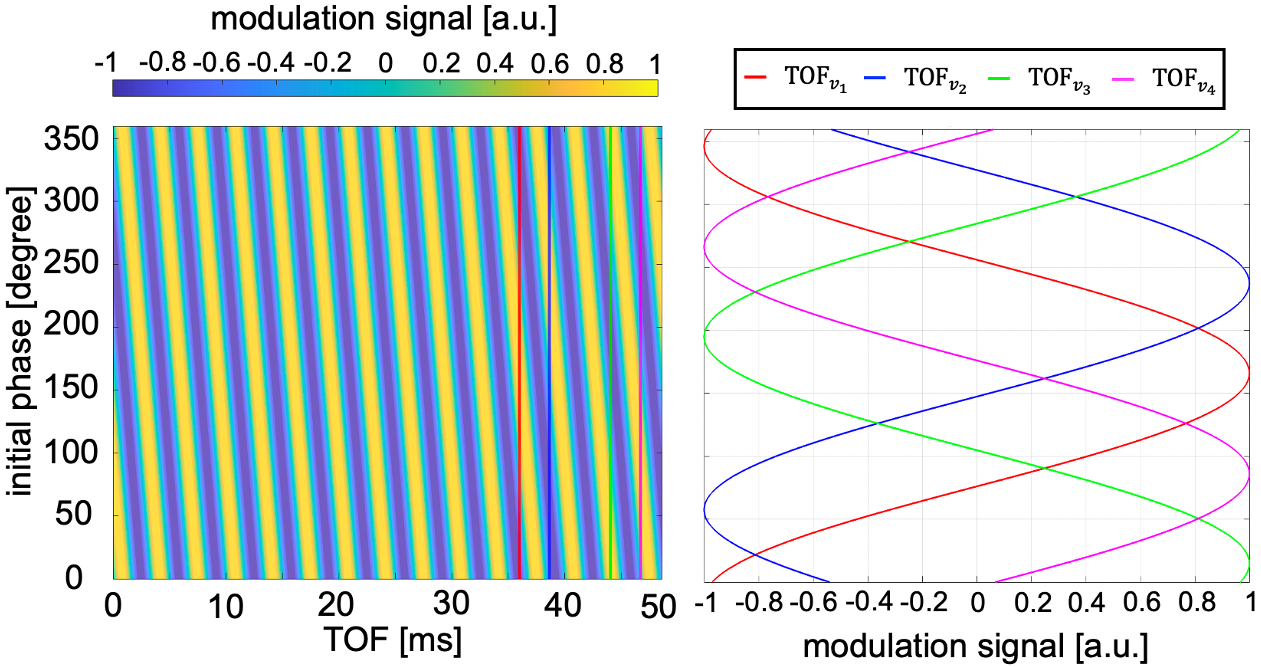}
   \caption{Simulated modulation signals for the $\mathrm{BS_L}$ noise at 300 Hz. All parameters and definitions in this figure are the same as those in Fig. \ref{fig:10}.}
   \label{fig:13}
\end{figure}

\noindent
The modulation signals as a function of the TOF of neutrons and the initial phase of the simulated signal, as well as those as a function of the initial phase at the TOF of each neutron for the five kinds of modulation signals, are shown in Figs. \ref{fig:10}-\ref{fig:13}.

\begin{figure}[H]
   \centering
   \includegraphics[clip,width=0.48\textwidth]{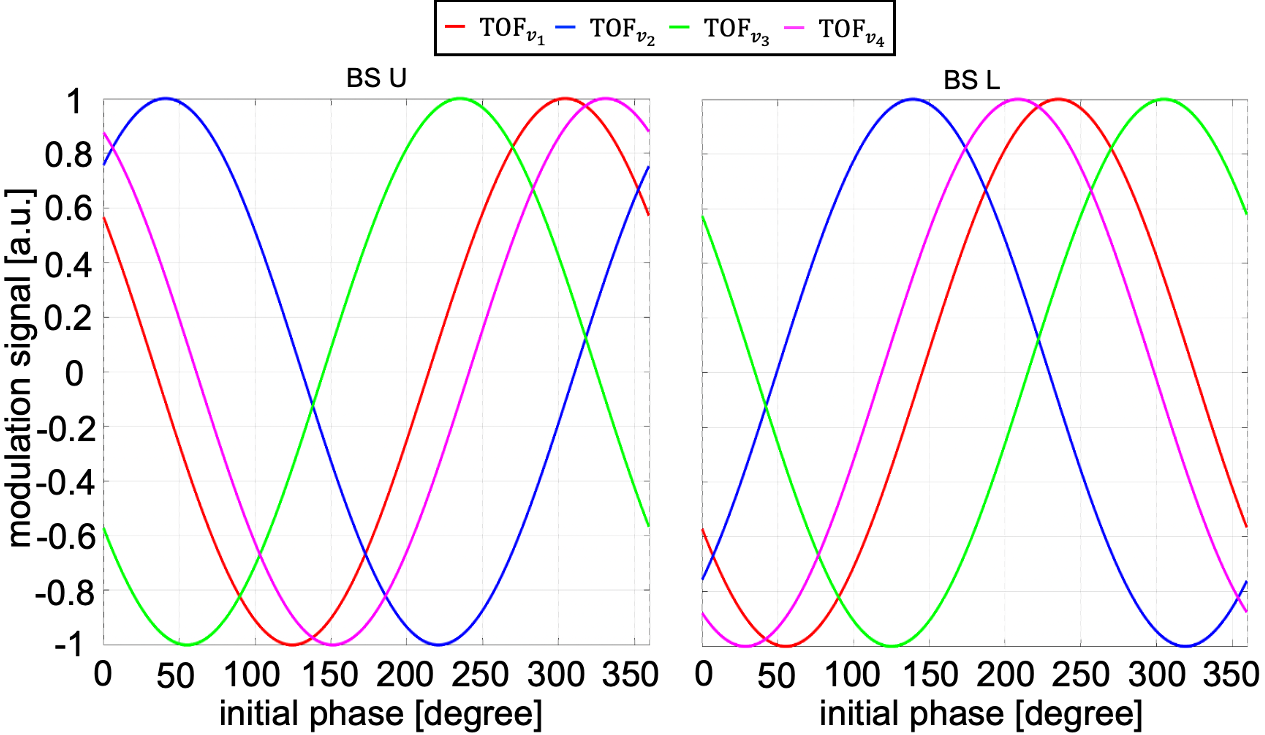}
   \caption{Simulated modulation signals as a function of the initial phase after compensation for $\mathrm{BS_U}$ (left panel) and $\mathrm{BS_L}$ (right panel) at the TOF of each neutron. Both panels share a common vertical axis.}
   \label{fig:14}
\end{figure}

\begin{figure}[H]
   \centering
   \includegraphics[clip,width=0.48\textwidth]{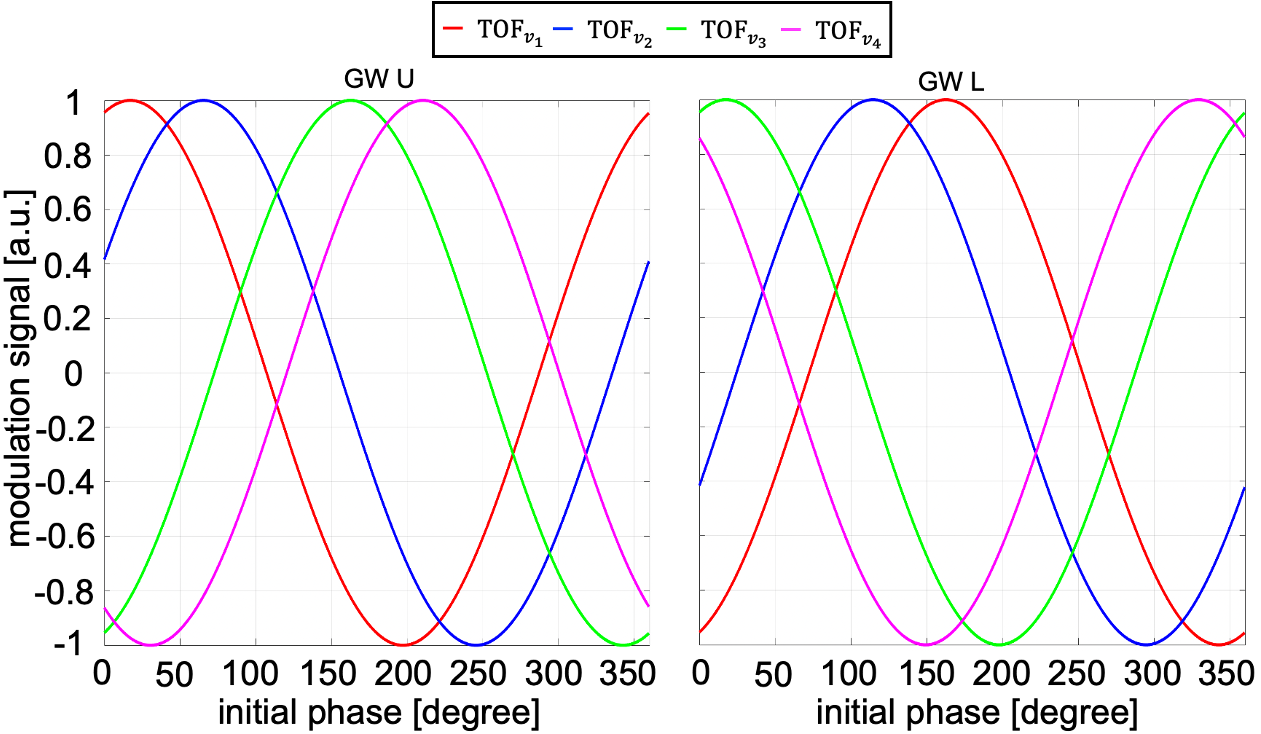}
   \caption{Simulated modulation signals as a function of the initial phase after compensation for $\mathrm{GW_U}$ (left panel) and $\mathrm{GW_L}$ (right panel) at the TOF of each neutron. All parameters and definitions in this figure are the same as those in Fig.  \ref{fig:14}.}
   \label{fig:15}
\end{figure}

Next, the compensation methods described in Fig. \ref{fig:9} are applied to the GW signals and the BS noises. Specifically, each curve in the right panel of Figs. \ref{fig:10}-\ref{fig:13} is shifted by an amount specified by the arrows in Fig. \ref{fig:9}. The modulation signals as a function of the initial phase after compensation are shown in Figs. \ref{fig:14}-\ref{fig:15}.

To explain this compensation mathematically, we provide an example of modulation signals affected by the $\mathrm{BS_U}$ modulation for each neutron speed as follows:

\begin{align}
   \phi_\mathrm{BS_U,1} (t) &= \sin \lbrace (2 \pi f_{\mathrm{M}} t_{v_1} + \phi_{\mathrm{com},v_1}) - 2 \pi \Delta f_{\mathrm{BS}} t_{v_1}  \rbrace \label{eq:11} , \\
   \phi_\mathrm{BS_U,2} (t) &= \sin \lbrace (2 \pi f_{\mathrm{M}} t_{v_2} + \phi_{\mathrm{com},v_2}) - 2 \pi \Delta f_{\mathrm{BS}} t_{v_2}  \rbrace \label{eq:12} , \\
   \phi_\mathrm{BS_U,3} (t) &= \sin \lbrace (2 \pi f_{\mathrm{M}} t_{v_3} + \phi_{\mathrm{com},v_3}) - 2 \pi \Delta f_{\mathrm{BS}} t_{v_3}  \rbrace \label{eq:13} , \\ 
   \phi_\mathrm{BS_U,4} (t) &= \sin \lbrace (2 \pi f_{\mathrm{M}} t_{v_4} + \phi_{\mathrm{com},v_4}) - 2 \pi \Delta f_{\mathrm{BS}} t_{v_4}  \rbrace \label{eq:14}.
\end{align}

\noindent
Here, $\phi_{\mathrm{com},v_\mathrm{i}}$ represents the phase compensation for each neutron speed. In Eqs. \eqref{eq:11}-\eqref{eq:14}, the first term of the sine function is adjusted by the phase compensation, while the second term represents the phase difference between each neutron speed. As a result, the relationship between the neutron phase changes received from a given simulated signal is consistent with the relationship shown in Table \ref{tab:1}. This approach accounts for the difference \ctext{4} (path length) in Table \ref{tab:2}.

\begin{figure}[H]
   \centering
   \includegraphics[clip,width=0.48\textwidth]{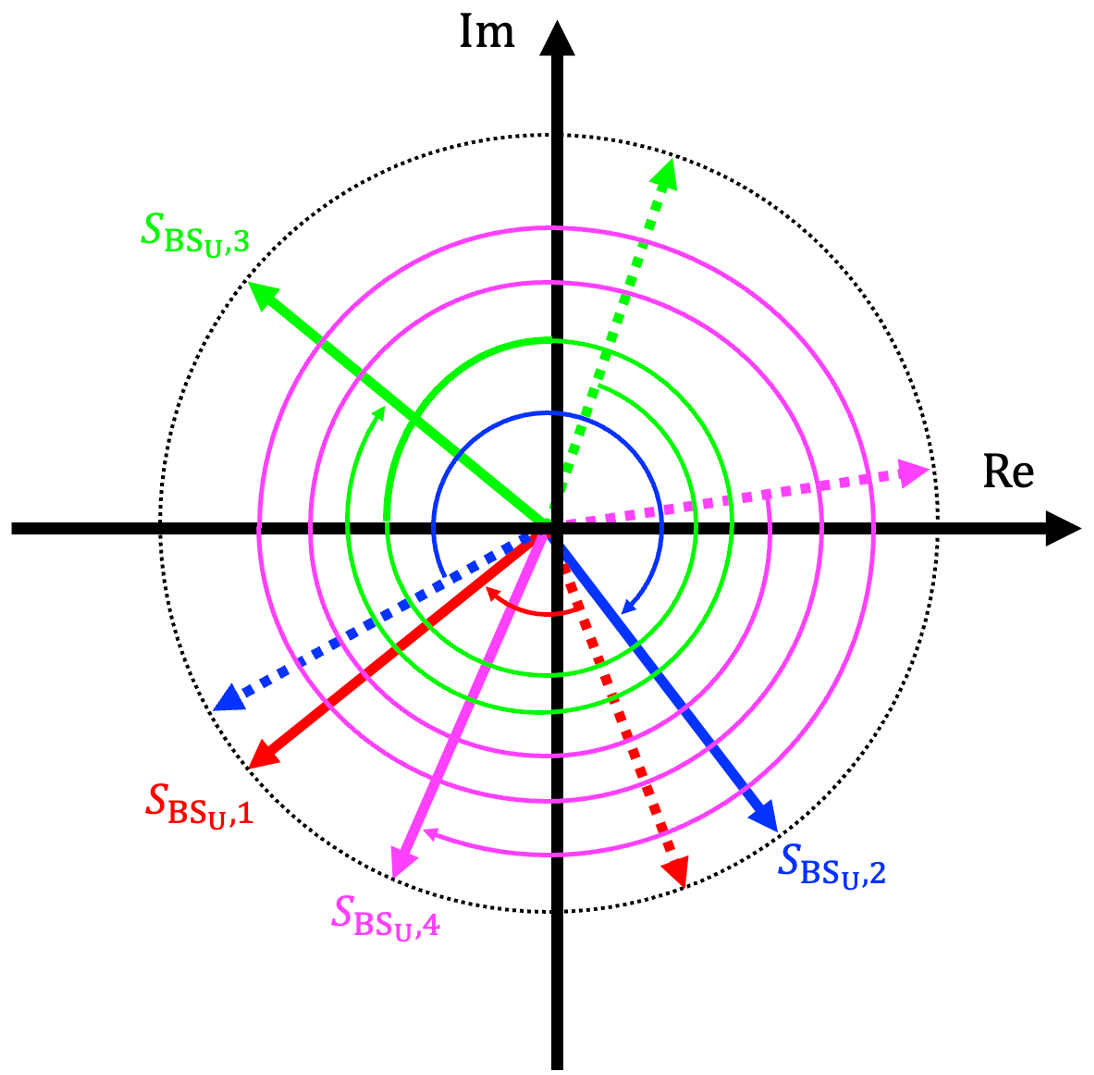}
   \caption{Simulated phasor diagram of $\mathrm{BS_U}$ displacement noises for the phase compensation. The dashed and solid arrows illustrate the BS displacement noises before and after phase compensation, respectively. Rotating arrows indicate phase compensation. The dashed circle shows the reference of normalized amplitude.}
   \label{fig:16}
\end{figure}

These compensations can be further explicitly explained for the $\mathrm{BS_U}$ case, as an example, using a phasor diagram, as shown in Fig. \ref{fig:16}. In this diagram, the dotted straight arrows representing the $\mathrm{BS_U}$ noises for each neutron speed are rotated by the rotating arrows corresponding to the phase compensation specified by the straight arrows in Fig. \ref{fig:9}. The resulting solid straight arrows illustrate the correct relationship between the four neutron speeds.

\begin{figure}[H]
   \centering
   \includegraphics[clip,width=0.4\textwidth]{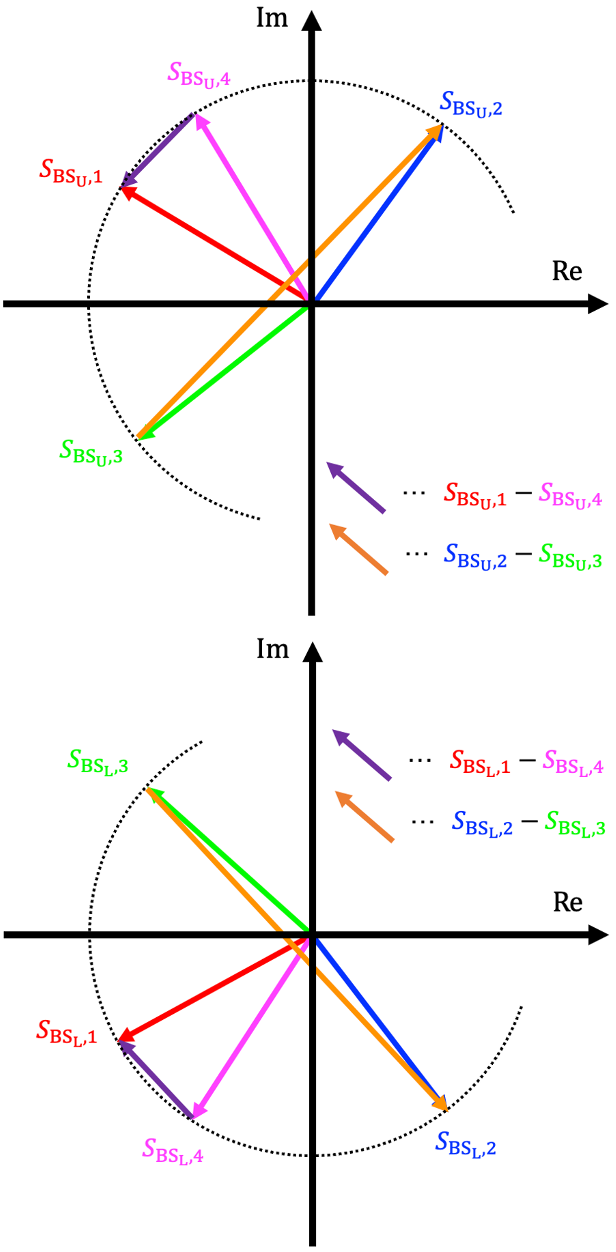}
   \caption{Simulated phasor diagram of $\mathrm{BS_U}$ (upper panel) and $\mathrm{BS_L}$ (lower panel) displacement noises, respectively. The red, blue, green, and magenta arrows show the neutron speeds of $v_1$, $v_2$, $v_3$, and $v_4$, respectively. The tangerine and purple arrows indicate the signals of Eqs. \eqref{eq:1} and \eqref{eq:2}. The dashed circle shows the reference of normalized amplitude.}
   \label{fig:17}
\end{figure}

\begin{figure}[H]
   \centering
   \includegraphics[clip,width=0.4\textwidth]{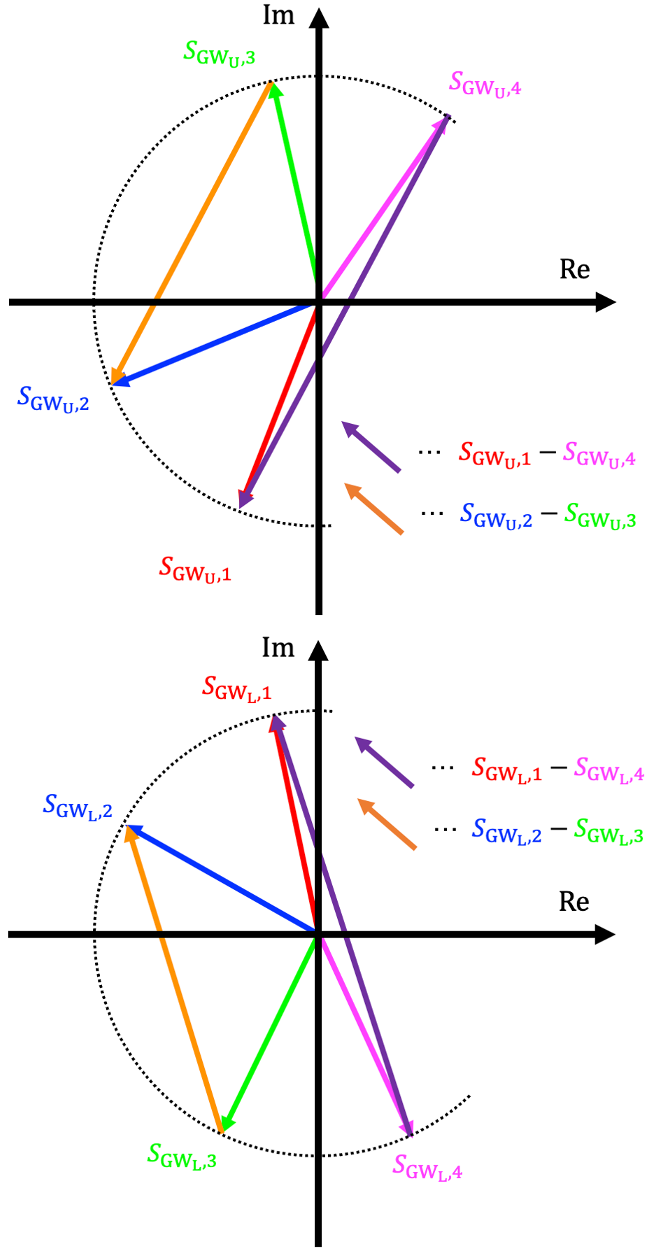}
   \caption{Simulated phasor diagram of $\mathrm{GW_U}$ (upper panel) and $\mathrm{GW_L}$ (lower panel) signals, respectively. All parameters and definitions in this figure are the same as those in Fig. \ref{fig:17}.}
   \label{fig:18}
\end{figure}

\noindent
After applying the same phase compensation, the phasor diagrams for $\mathrm{BS_U}$, $\mathrm{GW_U}$, $\mathrm{GW_L}$, and $\mathrm{BS_L}$, corresponding to Figs. \ref{fig:14}-\ref{fig:15} are shown in Figs. \ref{fig:17}-\ref{fig:18}. These figures also display the residual signals after all mirror noises are canceled as described in Eqs. \eqref{eq:1}-\eqref{eq:2}. It should be noted that the length ratio between the tangerine and purple arrows (which are parallel in each panel) for $\mathrm{BS_U}$ and $\mathrm{BS_L}$ is equal, and the same holds for $\mathrm{GW_U}$ and $\mathrm{GW_L}$. However, the length ratio for $\mathrm{BS_U}$ and $\mathrm{BS_L}$ differs from that for $\mathrm{GW_U}$ and $\mathrm{GW_L}$.

\begin{figure}[H]
   \centering
   \includegraphics[clip,width=0.48\textwidth]{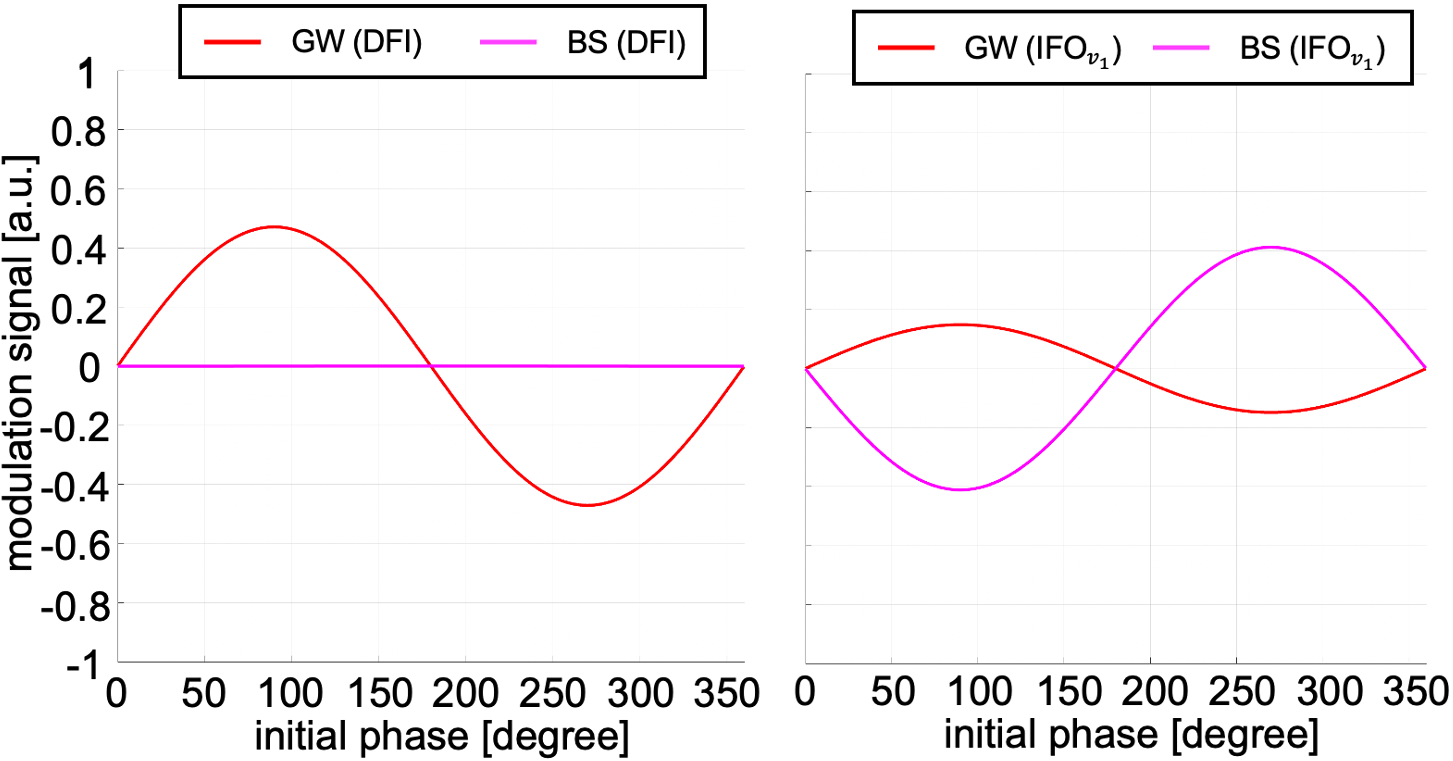}
   \caption{Simulated modulation signals for the BS noise (purple curve) and GW signal (red curve) as a function of the initial phase in the DFI combination signal (left panel) and in the interferometer signal of $v_1$ neutrons (right panel). Both panels share a common vertical axis.}
   \label{fig:19}
\end{figure}

Using the modulation signals after phase compensation, the DFI signal combination described in Eq. \eqref{eq:4} and the interferometer signal for $v_1$ neutrons (as an example) are shown in Fig. \ref{fig:19}.

\begin{figure}[H]
   \centering
   \includegraphics[clip,width=0.4\textwidth]{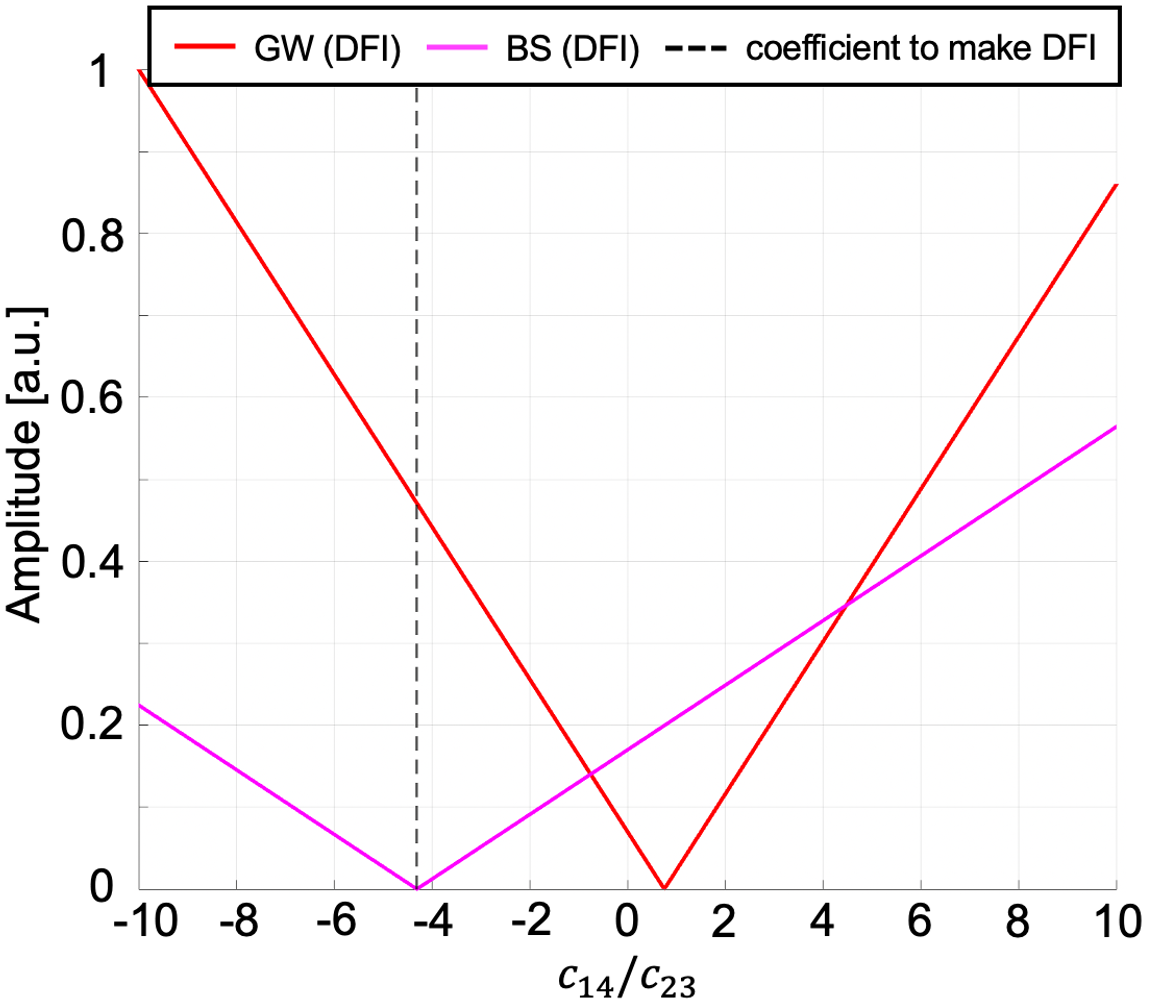}
   \caption{Simulated residuals of the GW signal (red curve) and BS noise (purple curve) in the DFI combination signal as a function of the coefficient ratio. The dashed line shows the correct DFI coefficient ratio.}
   \label{fig:20}
\end{figure}

 In the left panel of Fig. \ref{fig:19}, the BS noises (the purple curve) are completely canceled, while some GW signals (the red curve) remain. In the right panel of Fig. \ref{fig:19}, the interferometer signal for $v_1$ neutrons still contains BS noises. Here, the DFI coefficient ratio is $c_{14} / c_{23} \sim -4.31$. The dependence of the GW signals and BS noises in the DFI combination signal on the coefficient ratio is shown in Fig. \ref{fig:20}. 

This illustrates that a specific (correct) coefficient ratio that completely cancels the BS noise while preserving the GW signals.

The analysis methods in this section are summarized in Table \ref{tab:3}.

\begin{table}[h]
  \centering
  \caption{Solutions for addressing differences.}
  \scalebox{0.7}{
  {\renewcommand\arraystretch{2.0}
  \begin{tabular}{l||c|c} \hline
    & Solution & Category \\  \hline \hline
    I & Only BS noise is the cancellation target & \ctext{1} \\ \hline
    I\hspace{-1.2pt}I & GW polarization estimation & \ctext{2} \\ \hline
    I\hspace{-1.2pt}I\hspace{-1.2pt}I & Signal \& noise simulation by Al phase plate & \ctext{2} \\ \hline
    I\hspace{-1.2pt}V & Neutron count accumulation & \ctext{5} \\ \hline
    V  &  Mirror signal is used for time reference & \ctext{3} \\ \hline
    V\hspace{-1.2pt}I  & One pair of modulators to save space & \ctext{4} \\ \hline
    V\hspace{-1.2pt}I\hspace{-1.2pt}I  & One modulation per measurement & \ctext{4} \\ \hline
    V\hspace{-1.2pt}I\hspace{-1.2pt}I\hspace{-1.2pt}I  &  Frequency offset & \ctext{4} \\ \hline
  \end{tabular}}}
  \label{tab:3}
\end{table}

Using these analysis methods, the actual setup is a proof-of-principle experiment equivalent to the ideal setup.

\section{Experimental results}
\label{section:5}

This section presents the experimental results based on the analysis and experimental methods outlined in Section \ref{section:4} and summarized in Table \ref{tab:3}.

In this experiment, three distinct measurements were conducted: sample-out, sample-in without modulation, and sample-in with modulation, where "sample" refers to Al plates. The measurement sequence consisted of a sample-out measurement, followed by alternating sample-in measurements with modulation on and off. After completing eight pairs of sample-in measurements with different initial phases, the sample-out measurements were performed again. The phase scan was performed in steps of $\pi/4$. Each measurement requires approximately 10 minutes to accumulate neutron counts, resulting in an overall time of about three hours for measurements at a single modulation frequency.

\begin{figure}[H]
   \centering
   \includegraphics[clip,width=0.48\textwidth]{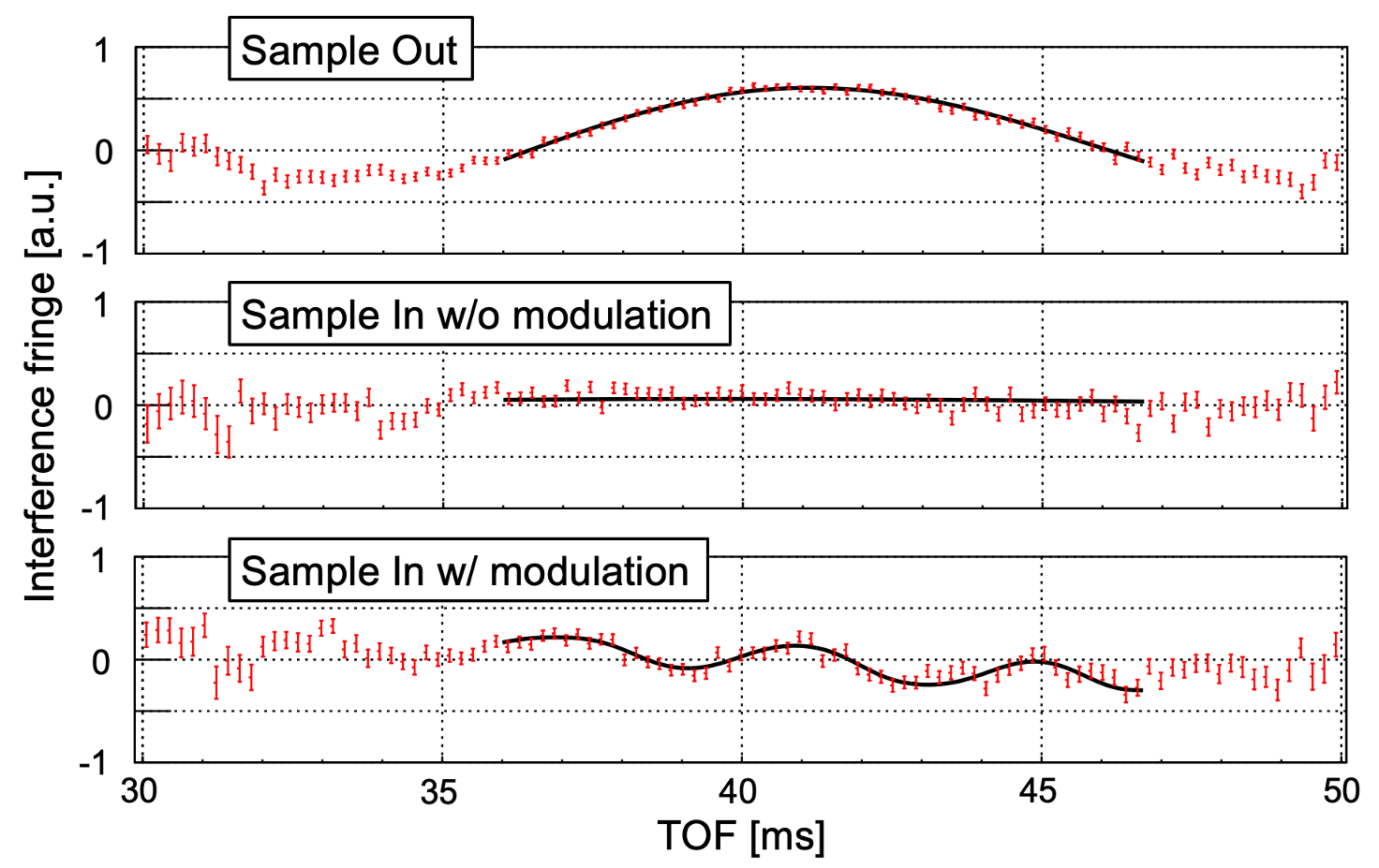}
   \caption{Measured interference fringes for different measurement types. The top panel corresponds to the interference fringe from a sample-out measurement, the middle panel to the sample-in measurement without modulation, and the bottom panel to the sample-in measurement with the 250 Hz modulation (as an example). These panels share a common horizontal axis. In these panels, the red markers represent the data points with associated error bars.}
   \label{fig:21}
\end{figure}

The three types of measurements for one phase scan are shown in Fig. \ref{fig:21}, using the measurement of the 250 Hz modulation as an example. It should also be noted that the path difference of the neutron interferometer was adjusted to be around the mid-fringe position, as seen in the middle and bottom panels of Fig. \ref{fig:21}.

These measurements are used to obtain the parameters of the modulation signals as follows. The interferometer parameters, such as the alignment error of the two etalons and the differences in SiO2 bases of the etalons in the two paths, are obtained from the sample-out measurement. The sample-in measurement provides the parameters of the modulators without modulation. The sample-in measurement with modulation provides the modulation parameters (called “AC”, hereafter) and the modulation offset. Considering these factors and the previous research \cite{Fujiie_PRL}, the fitting function can be defined as

\begin{align}
    I(\lambda) &= A \cos \Bigl \lbrace  \frac{P_L}{\lambda} + \Bigl( P_\mathrm{IFO} + P_\mathrm{sample} + P_\mathrm{offset} \notag \\    
    &+ P_\mathrm{AC} \sin \bigl( 2 \pi f_\mathrm{AC} t + \phi_\mathrm{initial} \bigr) \Bigr) \lambda \Bigr \rbrace + B,
    \label{eq:16}
\end{align}

\noindent
where $A$ is the visibility of the interference fringe, $B$ is the offset of the interference fringe, $P_L$ is the geometric optical length difference, $P_\mathrm{IFO}$ is the path difference from the $\mathrm{SiO_2}$ BSEs of the interferometer, $P_\mathrm{sample}$ is the path difference from a sample without modulation, $P_\mathrm{offset}$ is the offset of the path difference from a modulated sample, $P_\mathrm{AC}$ is the path difference from the modulation amplitude, $f_\mathrm{AC}$ is the modulation frequency, and $\phi_\mathrm{initial}$ is the initial phase of the modulation signal, which is used for the phase scan. The details of the fitting process are denoted in \ref{App}. The fitting results, including 100 Hz, 150 Hz, 250 Hz, and 300 Hz measurements, are shown in Table \ref{tab:4}.

\begin{table}[h]
   \centering
   \caption{Fitting parameter results for 100 Hz, 150 Hz, 250 Hz, and 300 Hz modulation.}
   \scalebox{1.0}{
   {\renewcommand\arraystretch{1.5}
   \begin{tabular}{c||c|c} \hline
     Frequency & $P_\mathrm{AC}$ $[\mathrm{nm}^{-1}]$& $f_\mathrm{AC}$ $[\mathrm{Hz}]$\\  \hline \hline
     100 Hz & $ 0.8589 \pm 0.0299 $ & $ 99.902 \pm 0.148 $ \\ \hline
     150 Hz & $ 0.8705 \pm 0.0323 $ & $ 149.611 \pm 0.164 $ \\ \hline
     250 Hz & $ 0.7217 \pm 0.0238 $ & $ 249.586 \pm 0.106 $ \\ \hline
     300 Hz & $ 0.7824 \pm 0.0260 $ & $ 299.576 \pm 0.102 $ \\ \hline
   \end{tabular}}}
   \label{tab:4}
 \end{table}

\begin{figure}[H]
   \centering
   \includegraphics[clip,width=0.48\textwidth]{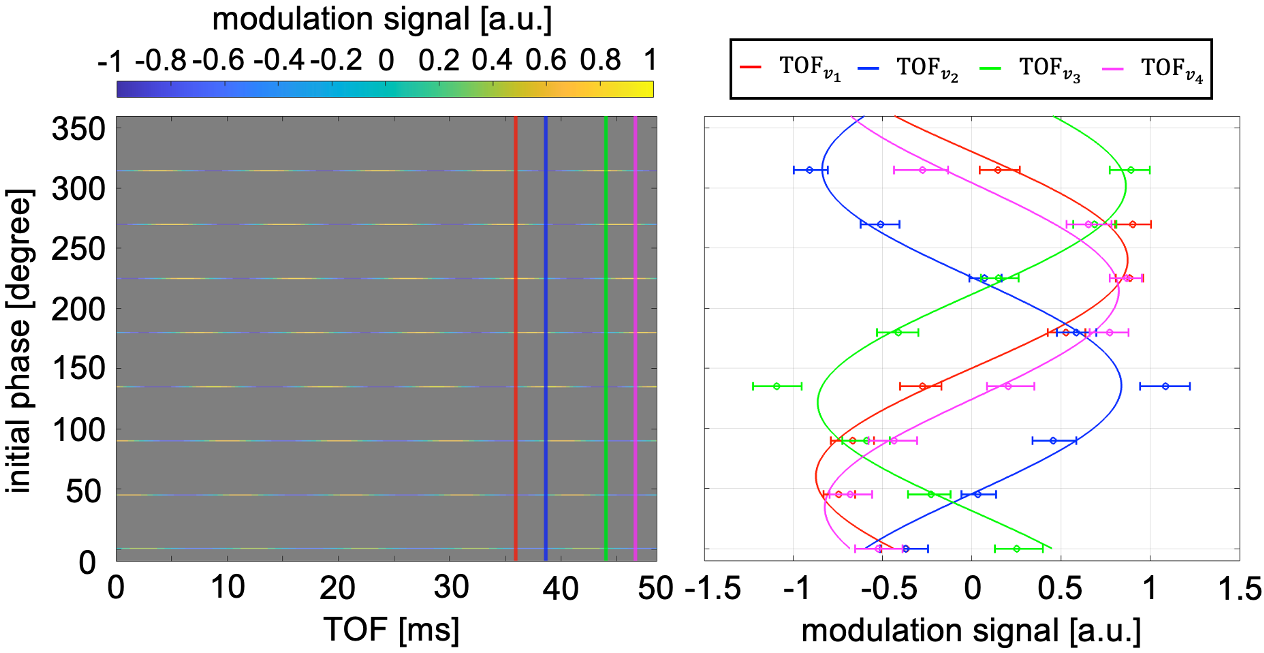}
   \caption{ Measured modulation signals for the $\mathrm{BS_U}$ noise at 100 Hz. The left panel shows the modulation signals as a function of the TOF of neutrons and the initial phase of the modulation signal. The right panel shows the modulation signals for neutrons of each speed as a function of the initial phase at the TOF corresponding to each neutron speed. The initial phase resolution is $\pi/4$. The data points and associated error bars are illustrated by dots and bars of various colors. The colored curves represent the fitting results. Both panels share a common vertical axis.}
   \label{fig:29}
\end{figure}

\begin{figure}[H]
   \centering
   \includegraphics[clip,width=0.48\textwidth]{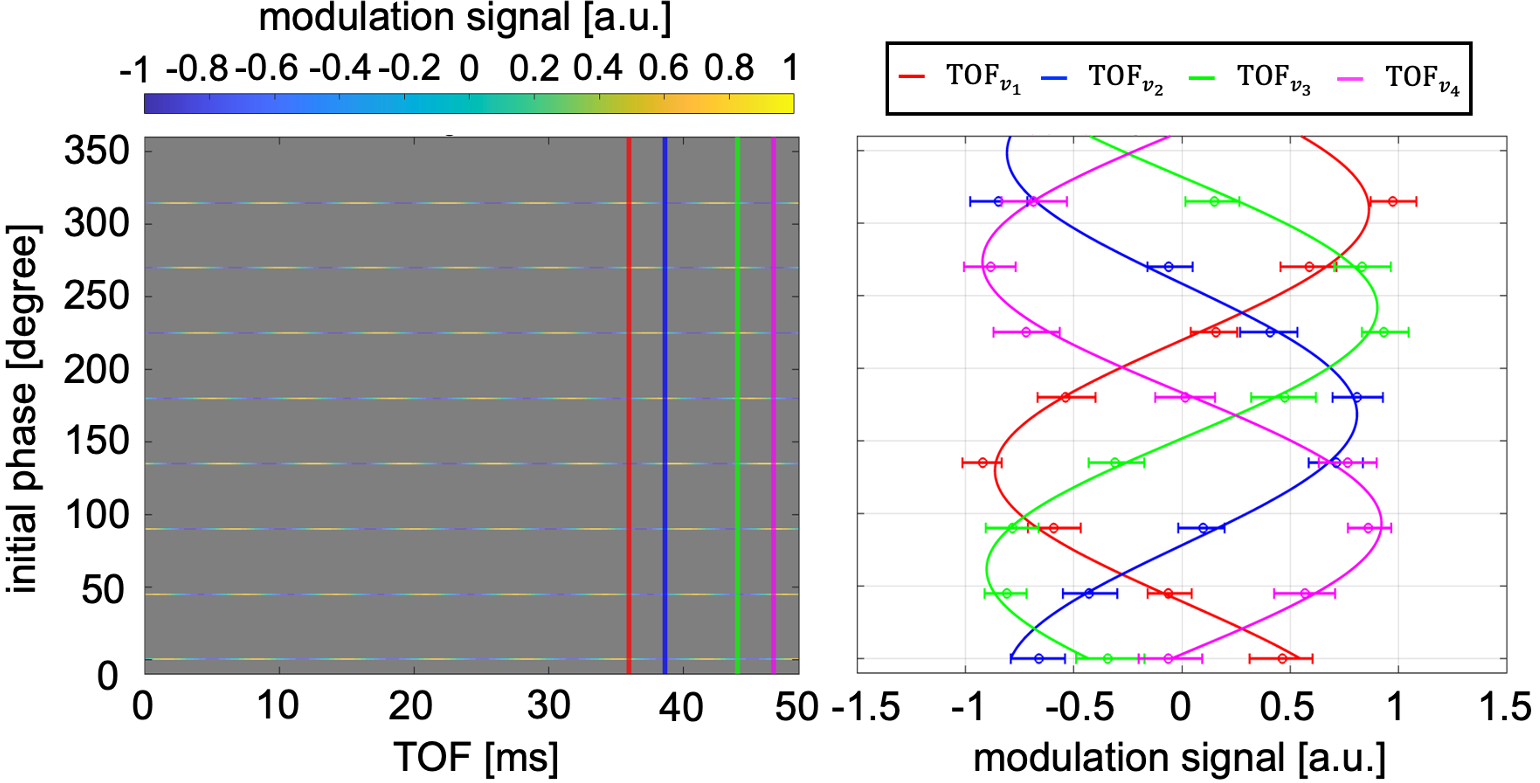}
   \caption{Measured modulation signals for the $\mathrm{GW_U}$ signal at 150 Hz. All parameters and definitions in this figure are the same as those in Fig. \ref{fig:29}.}
   \label{fig:30}
\end{figure}

\begin{figure}[H]
   \centering
   \includegraphics[clip,width=0.48\textwidth]{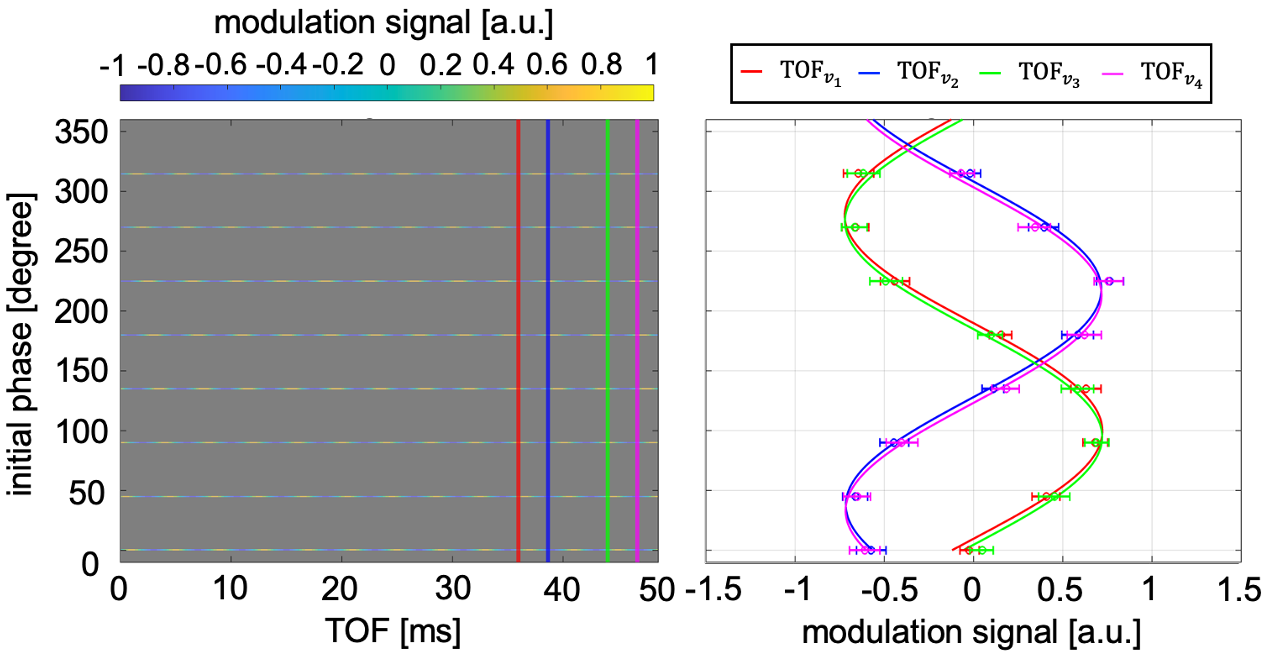}
   \caption{Measured modulation signals for the $\mathrm{GW_L}$ signal at 250 Hz. All parameters and definitions in this figure are the same as those in Fig. \ref{fig:29}.}
   \label{fig:31}
\end{figure}

\begin{figure}[H]
   \centering
   \includegraphics[clip,width=0.48\textwidth]{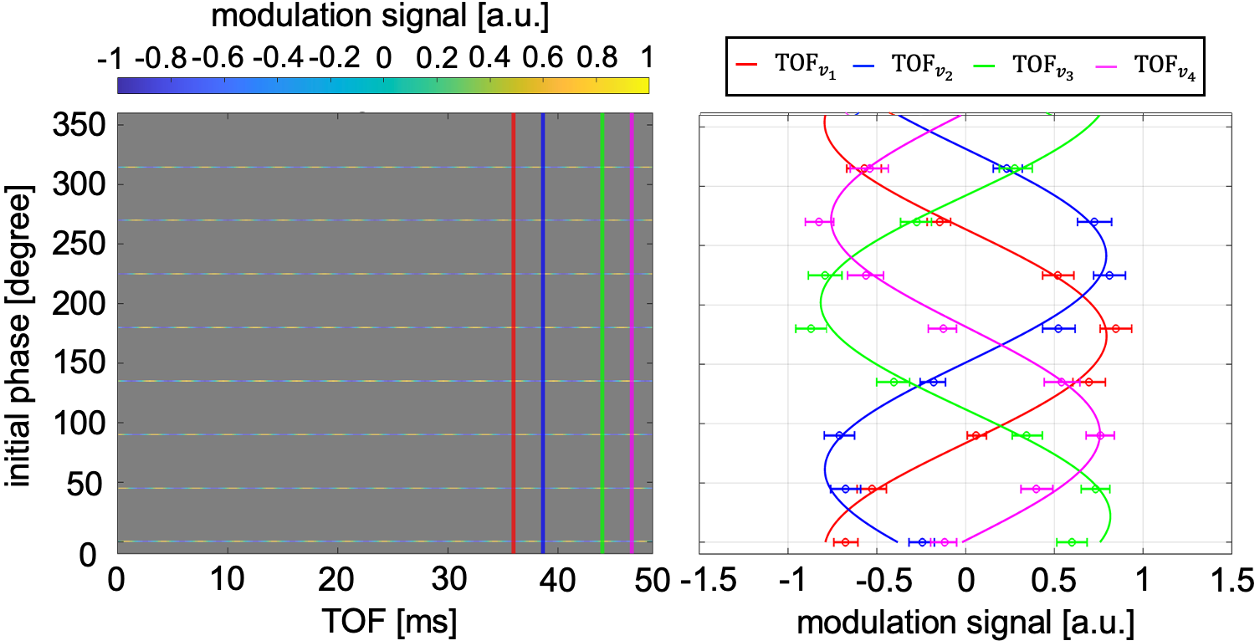}
   \caption{Measured modulation signals for the $\mathrm{BS_L}$ noise at 300 Hz. All parameters and definitions in this figure are the same as those in Fig. \ref{fig:29}.}
   \label{fig:32}
\end{figure}

Using these parameters, the modulation signals can be obtained through a similar process to that in Section \ref{subsection:2}. For the mirror noise modulation, the simulation data in Figs. \ref{fig:8}-\ref{fig:9} are used as a time reference. The interference fringes as a function of the TOF of neutrons and the initial phase of the simulated signal, as well as the interference fringes as a function of the TOF of neutrons, for $\mathrm{BS_U}$, $\mathrm{GW_U}$, $\mathrm{GW_L}$, and $\mathrm{BS_L}$ obtained from measurements are shown in Figs. \ref{fig:29}-\ref{fig:32}. Notably, the measured modulation signals agree with the simulated modulation signals shown in Figs. \ref{fig:10}-\ref{fig:13} within the margin of error, except for a slight difference in the overall magnitude.

\begin{figure}[H]
   \centering
   \includegraphics[clip,width=0.48\textwidth]{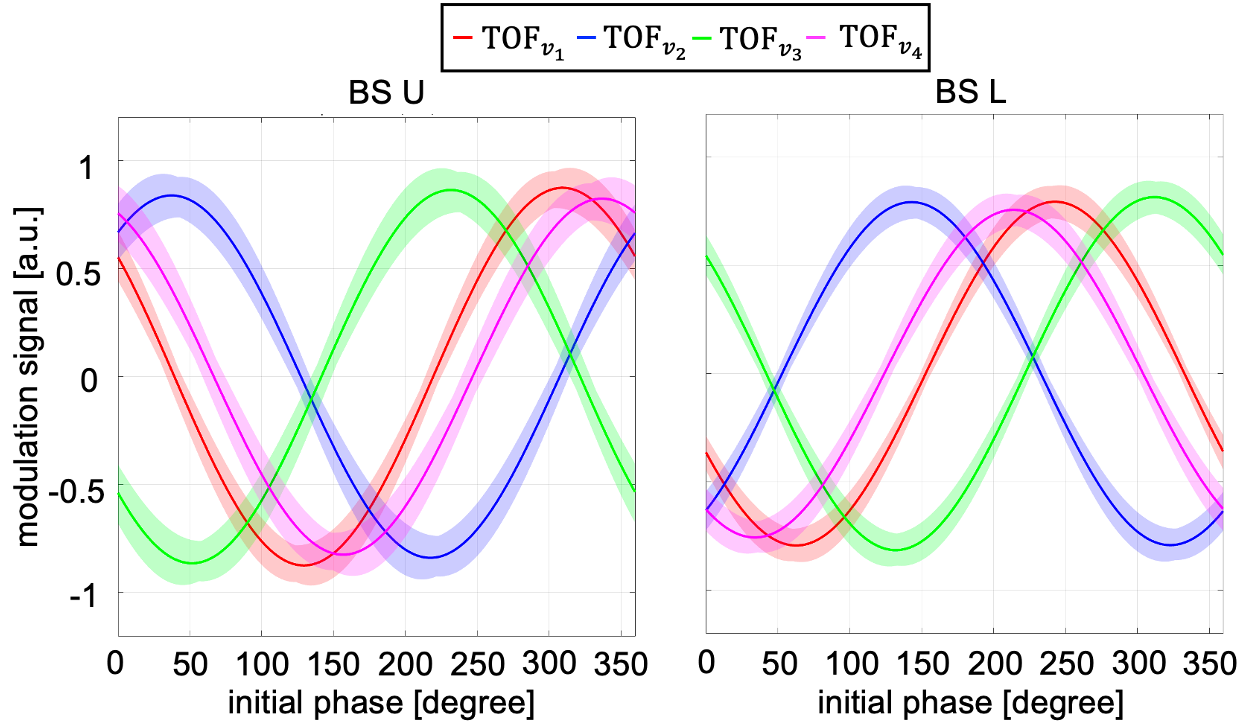}
   \caption{Measured modulation signals as a function of the initial phase after compensation for $\mathrm{BS_U}$ (left panel) and $\mathrm{BS_L}$ (right panel) at the TOF of each neutron. Both panels share a common vertical axis. The colored curves and shaded regions represent the fitting results and their associated errors.}
   \label{fig:33}
\end{figure}

\begin{figure}[H]
   \centering
   \includegraphics[clip,width=0.48\textwidth]{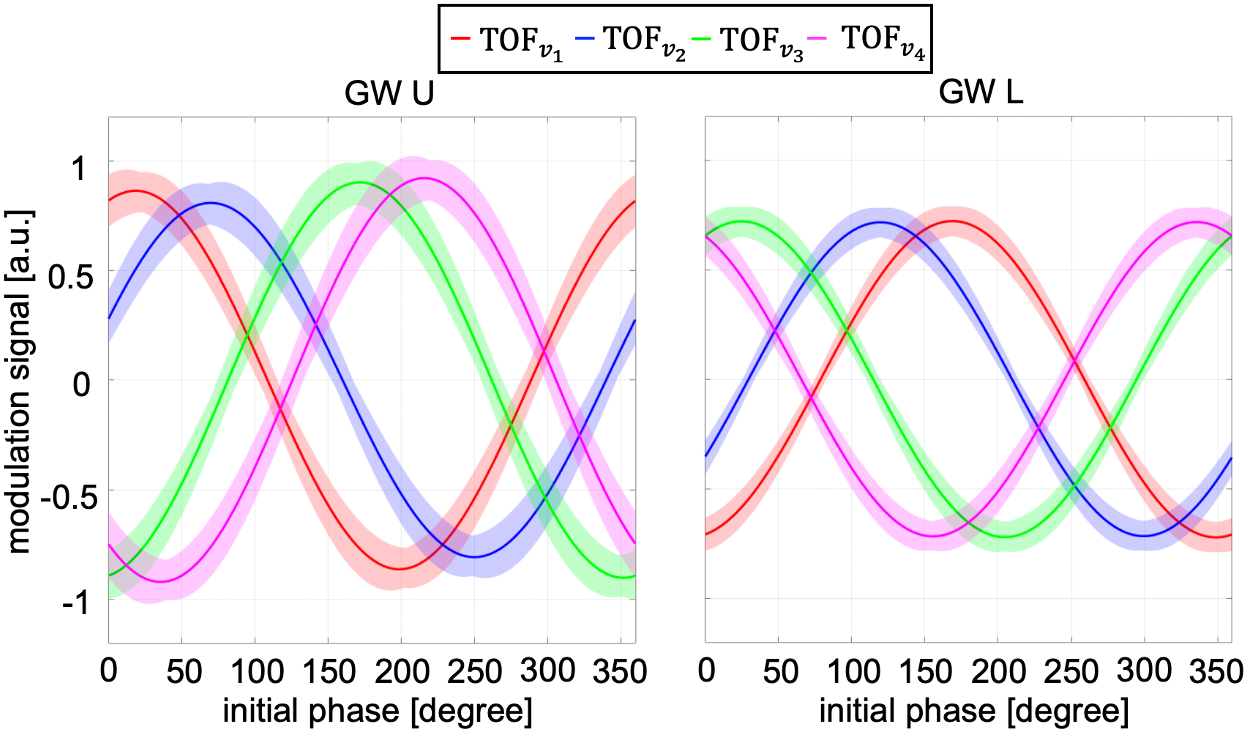}
   \caption{Measured modulation signals as a function of the initial phase after compensation for $\mathrm{GW_U}$ (left panel) and $\mathrm{GW_L}$ (right panel) at the TOF of each neutron. All parameters and definitions in this figure are the same as those in Fig. \ref{fig:33}}
   \label{fig:34}
\end{figure}

In Figs. \ref{fig:29}-\ref{fig:32}, the curves representing modulation signals are obtained by fitting the data with a sine function. Subsequently, the same compensation method as in Fig. \ref{fig:9} is applied to the GW signals and the BS noises, as shown in Figs. \ref{fig:33}-\ref{fig:34}. The measured modulation signals after compensation also agree with the simulated modulation signals shown in Figs. \ref{fig:14}-\ref{fig:15} within the margin of error, except for a slight difference in the overall magnitude.

\noindent
Figs. \ref{fig:35}-\ref{fig:36} show the measured phasor diagrams corresponding to the $\mathrm{BS_U}$, $\mathrm{BS_L}$, $\mathrm{GW_U}$, and $\mathrm{GW_L}$ signals. These diagrams also illustrate the residual signals after cancellation of all mirror noises. The measured phasor diagram agrees with the simulated phasor diagram shown in Figs. \ref{fig:17}-\ref{fig:18}, including the relationships between the arrows, within the margin of error, except for a slight difference in the overall magnitude.

\begin{center}
   \includegraphics[clip,width=0.4\textwidth]{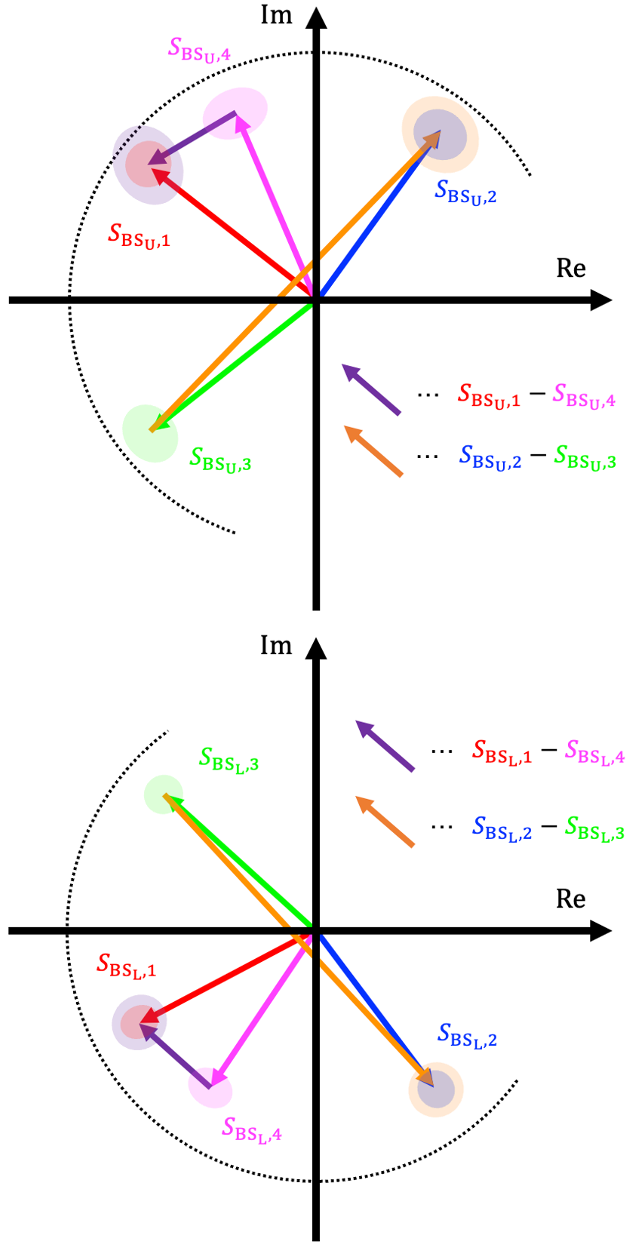}
   \captionof{figure}{Measured phasor diagram of measured $\mathrm{BS_U}$ (upper panel) and $\mathrm{BS_L}$ (lower panel) displacement noises, respectively. The red, blue, green, and magenta arrows show the neutron speeds of $v_1$, $v_2$, $v_3$, and $v_4$, respectively. The tangerine and purple arrows indicate the signals defined in Eqs. \eqref{eq:1} and \eqref{eq:2}. The shaded regions represent their errors in amplitude and phase, which are colored according to their associated arrows. The dashed circle indicates the amplitude of the simulation result.} % capt-of パッケージを使用
   \label{fig:35}
\end{center}

\begin{center}
   \includegraphics[clip,width=0.4\textwidth]{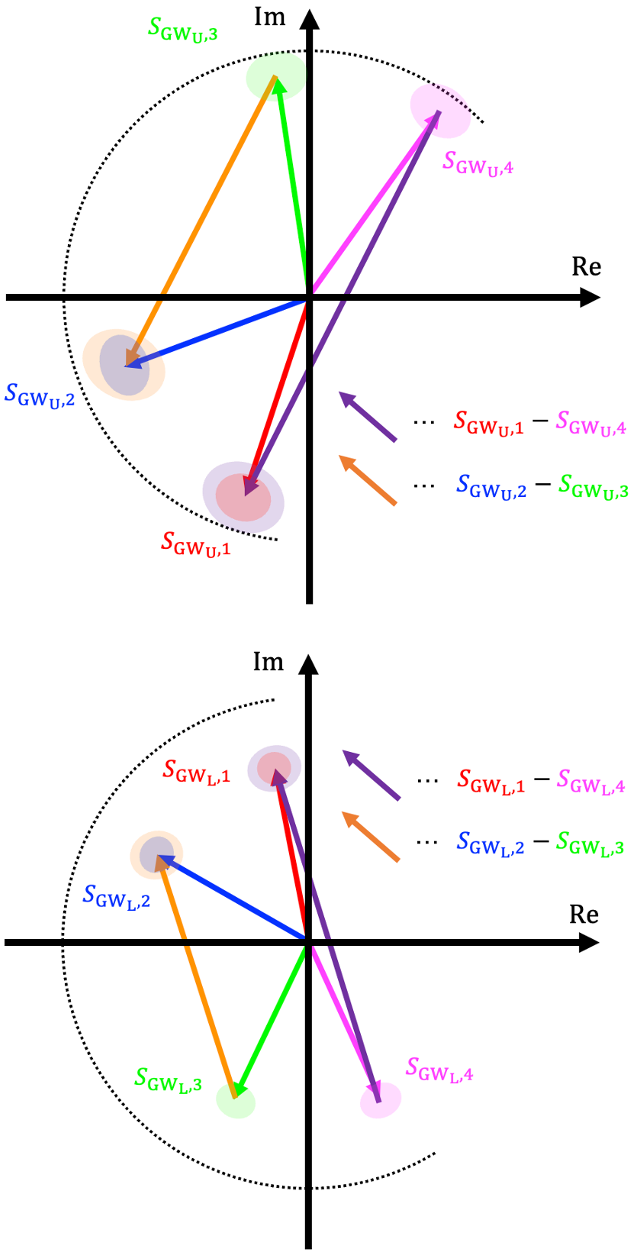}
   \captionof{figure}{Measured phasor diagram of measured $\mathrm{GW_U}$ (upper panel) and $\mathrm{GW_L}$ (lower panel) signals. All parameters and definitions in this figure are the same as those in Fig. \ref{fig:35}.}
   \label{fig:36}
\end{center}

\begin{figure}[H]
   \centering
   \includegraphics[clip,width=0.4\textwidth]{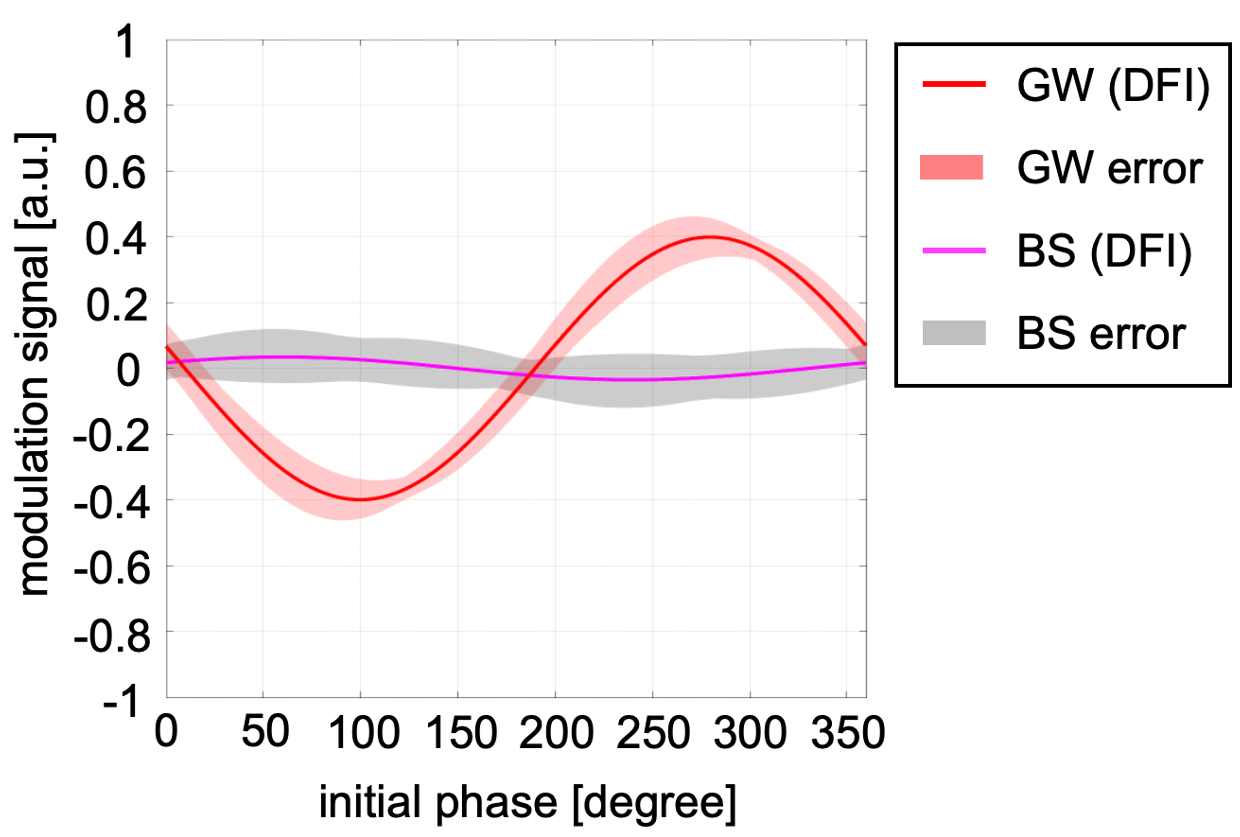}
   \caption{Measured modulation signals for the BS noise (purple curve) and GW signal (red curve) as a function of the initial phase in the DFI combination signal. The colored curves and shaded regions represent the fitting results and their associated errors}
   \label{fig:37}
\end{figure}

The measured modulation signals for the BS noise and GW signal as a function of the initial phase in the DFI combination signal are shown in Fig. \ref{fig:37}. The BS noises are effectively canceled, while preserving some GW signals. This also agrees with the simulated result in the left panel of Fig. \ref{fig:19}, within the margin of error.

Fig. \ref{fig:37} shows the experimental result demonstrating the dependence of the GW signals and BS noises in the DFI combination signal on the coefficient ratio is shown, along with the error regions. The shaded error regions represent 1$\sigma$ fitting uncertainties obtained from sine-wave fits to the data, incorporating counting-statistic uncertainties from neutron detection, as shown in Figs. \ref{fig:29}–\ref{fig:32} and \ref{App}. This result shows that BS noises are canceled at the correct coefficient ratio within the margin of error, while preserving the GW signal.

\begin{figure}[H]
   \centering
   \includegraphics[clip,width=0.4\textwidth]{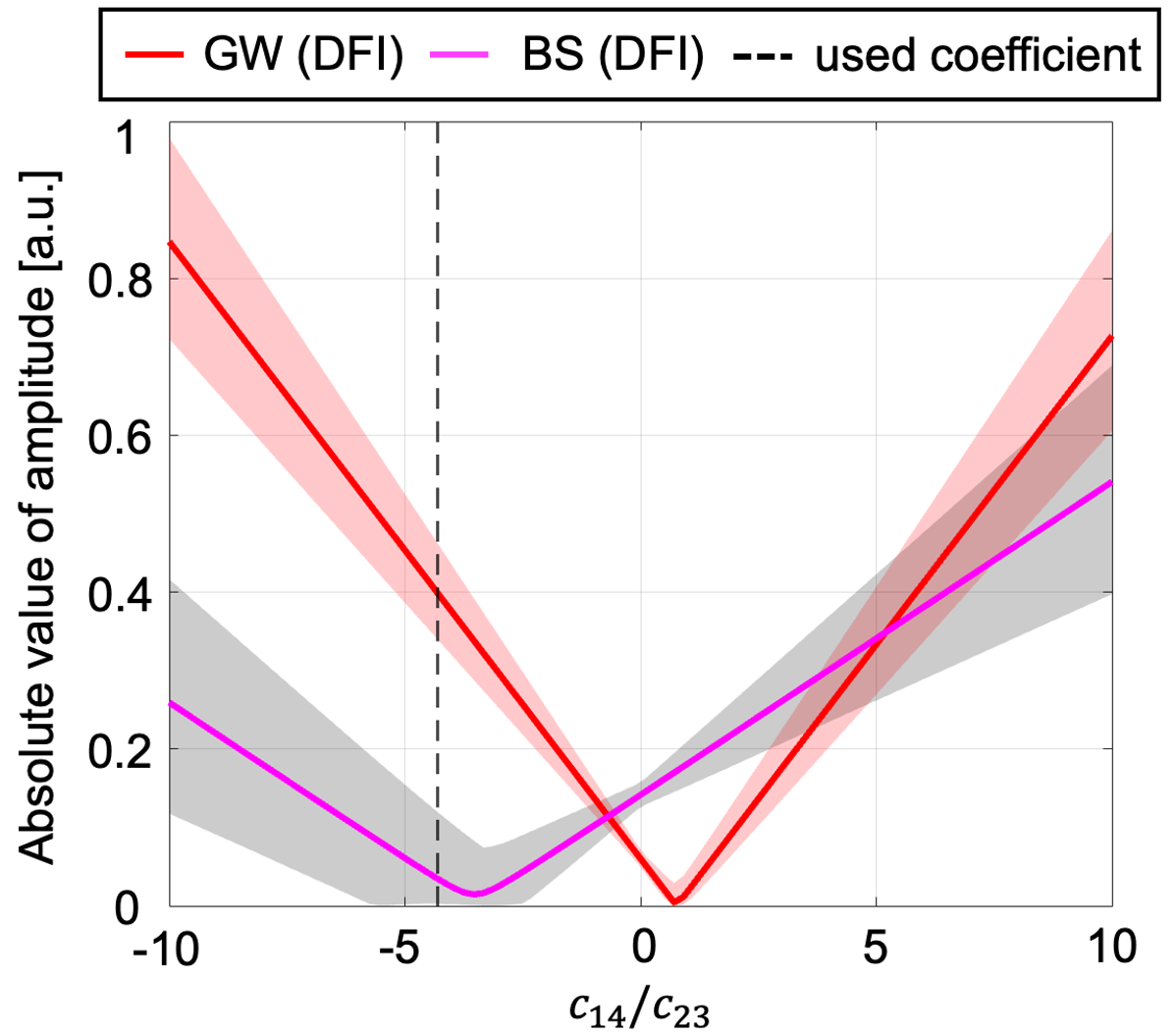}
   \caption{Measured residuals of the GW signal (red curve) and BS noise (purple curve) in the DFI combination signal as a function of the coefficient ratio. The shaded regions represent the associated errors. The dashed line indicates the correct DFI coefficient ratio obtained from the simulation.}
   \label{fig:38}
\end{figure}

Fig. \ref{fig:38} shows the dependence of the BS noise and GW signal in the DFI combination signal on the coefficient ratio is shown, along with the error regions. The definition of shaded error bands in Fig. \ref{fig:38} is the same as that in Fig. \ref{fig:37}. The figure shows that the coefficient ratio that cancels out the BS noise matches the predicted value ($c_14/c_23 \sim -4.31$) from the simulation within the margin of error. Importantly, despite the errors, the GW signal is unequivocally preserved at this correct coefficient ratio.

\section{Conclusion}
\label{section:6}

We successfully completed a proof-of-principle neutron DFI experiment using the pulsed neutron source at J-PARC. This experiment marks the first demonstration of neutron DFI and of a neutron interferometer for GW detection. While this experiment is based on the theoretical framework established in previous research \cite{DFNI_Iwaguchi_2}, five significant differences arose between the ideal and actual experimental setups due to various constraints. We optimized the experimental setup to address these differences and developed a novel analytical method tailored for the proof-of-principle experiment. The experimental results of the DFI signal combination show that the BS noise in the DFI signal is effectively canceled, while the GW signal is preserved. This experiment marks the first demonstration of both neutron DFI and a neutron interferometer for GW detection.

It should be noted that there are challenges in achieving the sensitivity required to detect GWs, as mentioned in the previous research \cite{DFNI_Nishizawa}. While a practical neutron GW detector remains far beyond the reach of current technology, future advances in neutron optics and sources may eventually make such a detector feasible. Importantly, the data analysis methods introduced in this paper enable us to conduct experiments for a neutron GW detector without waiting for advances in neutron optics technology, even though current neutron interferometry has not yet reached the level of laser-based systems and is still under development for GW detection.

Our results show good agreement between measured and simulated modulation signals, validating the experimental strategy and the analysis techniques. These findings suggest that neutron DFI is a promising candidate for future GW detection and other precision measurements.

\section{Acknowledgments}
We thank Kushal Jain for assistance with English editing. This work was supported by the Japan Society for the Promotion of Science (JSPS) KAKENHI Grant Number JP19K21875. The neutron experiment at the Materials and Life Science Experimental Facility of J-PARC was performed under  S-type projects of KEK IMSS (Proposal No. 2019S03). A. N. is supported by JSPS KAKENHI Grants No. JP23K03408, No. JP23H00110, and No. JP23H04893. Y. M. is supported by JSPS KAKENHI Grant Nos. JP20H05850, JP20H05854, JP20H05639, JP24K21546 and JP24K00640.

\renewcommand{\thesection}{Appendix~\Alph{section}}
\setcounter{section}{0} % Appendix A から始める
\section{Fitting process for experimental results}
\label{App}

In this appendix, we provide the experimental fitting process and its results, which are obtained by the fitting function written in Eq. \eqref{eq:16}. Before using this fitting function, the parameters of A and B with the sample need to be determined by the yaw-scan measurement. This is to measure the interference when the BSE is rotated in the yaw direction relative to the neutron beam. Specifically, there are three types of fitting functions, which are written as

\begin{align}
   I_0(\lambda) &= A_0 \cos \Bigl \lbrace  \frac{P_{L}}{\lambda} + P_\mathrm{IFO} \lambda \Bigr \rbrace + B_0, \\
   I_1(\lambda) &= A_\mathrm{fixed} \cos \Bigl \lbrace \frac{P_{L}}{\lambda} + \Bigl( P_\mathrm{IFO} + P_\mathrm{sample} \Bigr) \lambda \Bigr \rbrace + B_\mathrm{fixed}, \\
   I_2(\lambda) &= A_\mathrm{fixed} \cos \Bigl \lbrace  \frac{P_{L}}{\lambda} + \Bigl( P_\mathrm{IFO} + P_\mathrm{sample} + P_\mathrm{offset} \notag \\
   & + P_\mathrm{AC} \sin \bigl( 2 \pi f_\mathrm{AC} t + \phi_\mathrm{initial} \bigr) \Bigr) \lambda \Bigr \rbrace + B_\mathrm{fixed}.
\end{align}

\noindent
where $I_0(\lambda)$ is used for sample-out measurements to determine $P_{L}$ and $P_\mathrm{IFO}$, $I_1(\lambda)$ is used for sample-in measurements without modulation to get $P_\mathrm{sample}$, and $I_2(\lambda)$ is used for sample-in measurements with modulation to obtain $P_\mathrm{offset}$, $P_\mathrm{AC}$ and $f_\mathrm{AC}$. Before each frequency measurement, the yaw angle of the BSE, which maximizes the visibility of the interference fringe, is determined by the yaw-scan measurement. The alignment of the stage position can change the yaw angle of the BSE. The results of yaw-scan measurements are shown in Figs. \ref{fig:Ap-1} and \ref{fig:Ap-2}.

\begin{center}
   \includegraphics[clip,width=0.48\textwidth]{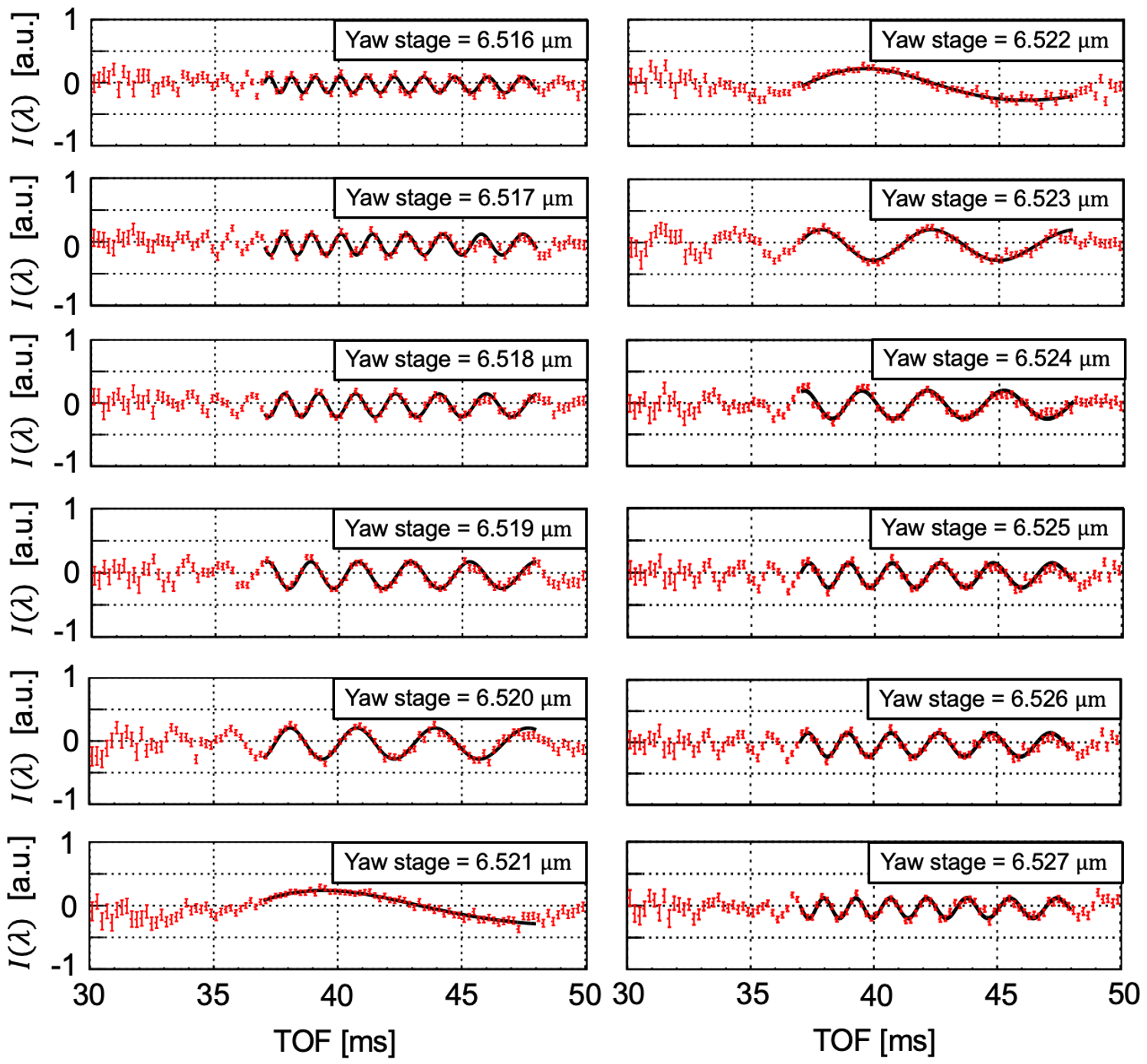}
   \captionof{figure}{Interference fringes from yaw-scan measurements. In these panels, the red markers represent the data points with error bars, and the black curves are the fitting curves. Each measurement corresponds to a specific position of the yaw stage at the BSE, as shown in each panel.} % capt-of パッケージを使用
   \label{fig:Ap-1}
\end{center}

\begin{center}
   \includegraphics[clip,width=0.48\textwidth]{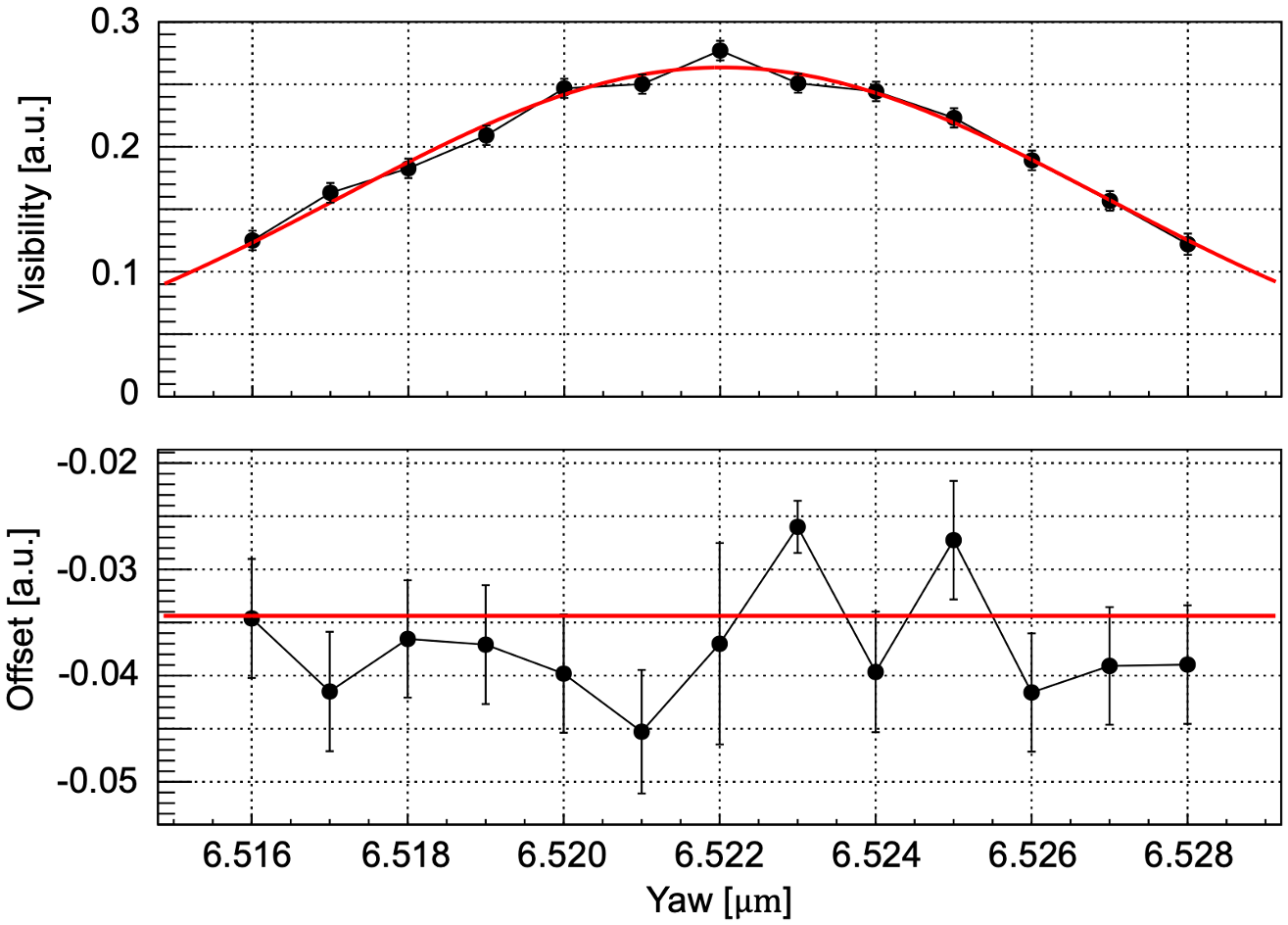}
   \captionof{figure}{Fitting results from yaw-scan measurements. In both panels, the black lines represent the data points with error bars, and the red lines indicate the fitting curves. The upper and lower panels show the visibility $A_1$, the offset $B_1$ and the corresponding fitting curves when a step of 1 $\mu$m changes the yaw stage position. A Gaussian curve fits the visibility, and a constant fits the offset.} % capt-of パッケージを使用
   \label{fig:Ap-2}
\end{center}

\noindent
These results indicate that $A_\mathrm{fixed} = (0.2635\pm0.0036)$, $B_\mathrm{fixed} = (0.0344\pm0.0014)$, and that the yaw stage position that maximizes the visibility is 6.522 $\mathrm{\mu m}$. In other sample-in measurements, the position of the BSE was set to the yaw angle determined here, and the visibility of $A_1$ and the offset of $B_1$ obtained from the yaw-scan measurements were fixed.

\begin{figure}
   \centering
   \includegraphics[clip,width=0.45\textwidth]{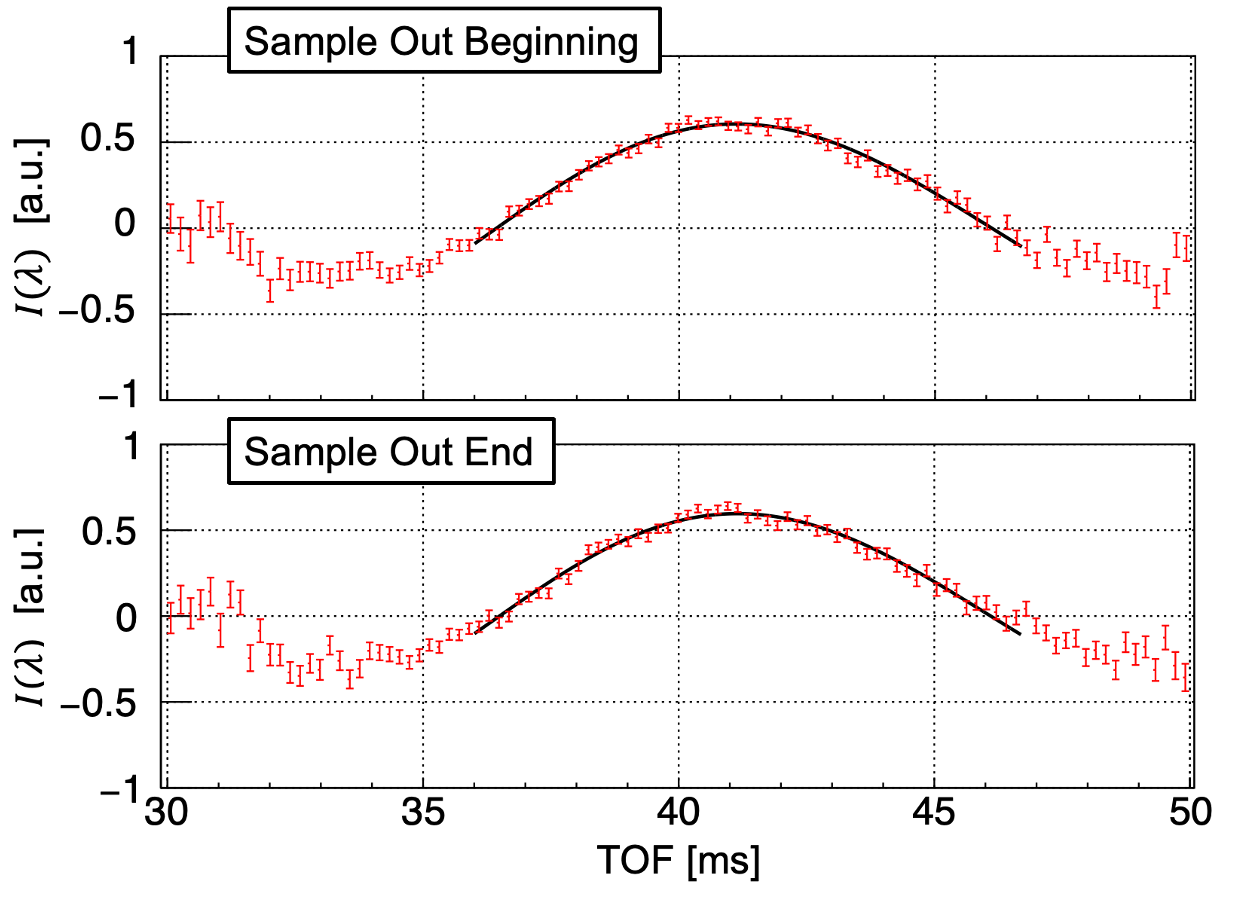}
   \caption{Interference fringes from the sample-out measurements for the 250 Hz modulation. The upper and lower panels indicate the fringes at the beginning and end of all the measurements, respectively. In these panels, the red markers represent the data points and error bars, while the black curves are the fitting curves.} % capt-of パッケージを使用
   \label{fig:22}
\end{figure}

For the DFI signal combination, it is necessary to determine the modulation parameters of $P_\mathrm{AC}$ and $f_\mathrm{AC}$. The process for obtaining these parameters is described below, using data from the 250 Hz measurement as an example. The fitting range for all processes is set to 36–46.5 ms in TOF, corresponding to the reflection bandwidth of the neutron mirror. As a first step, $P_L$ and $P_\mathrm{IFO}$ are determined through fitting based on the sample-out measurements, which are taken at the beginning and end of the measurements to eliminate the effects of interferometric disturbances. These results are shown in Fig. \ref{fig:22}.

This setup had already been adjusted to maintain the mid-fringes before the sample-in measurements, based on the yaw-scan measurements. The fitting results for $P_L$ and $P_\mathrm{IFO}$ from these sample-out measurements are shown in Fig. \ref{fig:23}.

% \begin{table}[H]
%    \centering
%    \begin{minipage}{0.23\textwidth}
%        \centering
%        \includegraphics[clip,width=\textwidth]{p0out_250Hz_fit.png}
%    \end{minipage}
%    \begin{minipage}{0.237\textwidth}
%        \centering
%        \includegraphics[clip,width=\textwidth]{p2out_250Hz_fit.png}

%    \end{minipage}
%    \captionof{figure}{Fitting results of $P_L$ and $P_\mathrm{IFO}$. The black markers represent data points with error bars, while the red lines show the fitting curves. The horizontal axis corresponds to the measurement order. In the left panel, the fitting function is a linear. In the right panel, the fitting function is a constant.}
%    \label{fig:23}
% \end{table}

\begin{table}[H]
   \centering
   \begin{minipage}{0.45\textwidth}
       \centering
       \includegraphics[clip,width=\textwidth]{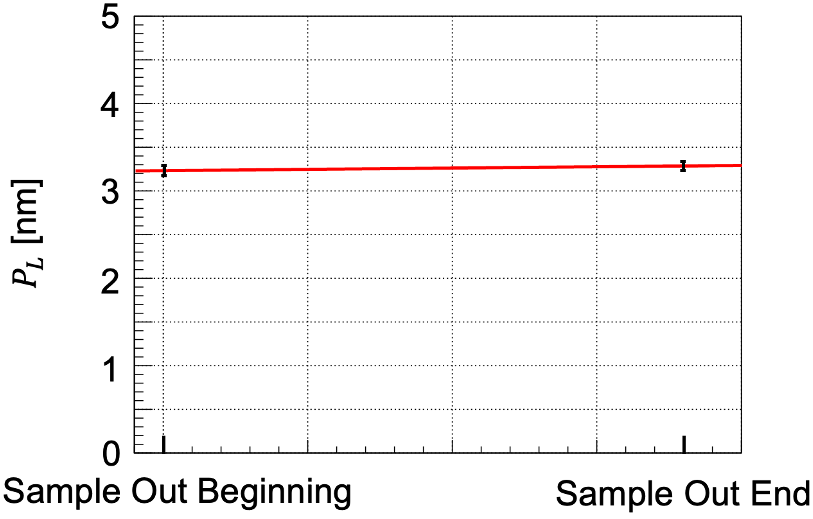}
   \end{minipage}

   \vspace{1em} % 図の間隔を調整

   \begin{minipage}{0.45\textwidth}
       \centering
       \includegraphics[clip,width=\textwidth]{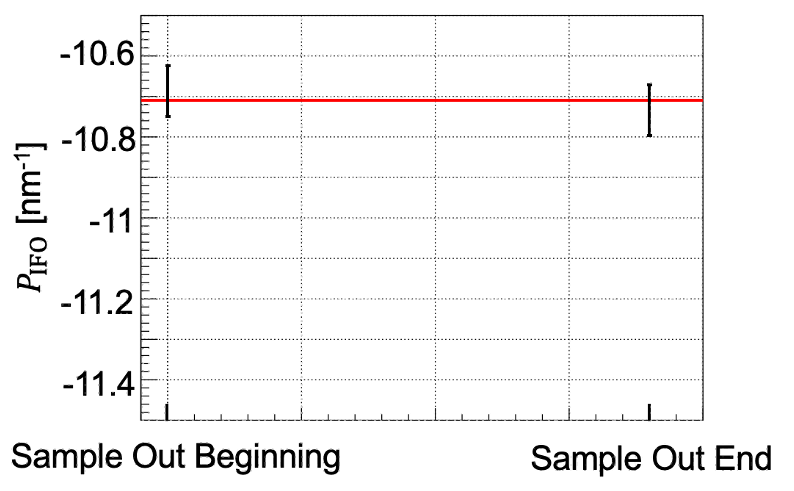}
   \end{minipage}

   \captionof{figure}{Fitting results of $P_L$ and $P_\mathrm{IFO}$. 
   The black markers represent data points with error bars, while the red lines show the fitting curves. 
   The horizontal axis corresponds to the measurement order. 
   In the top panel, the fitting function is a linear. 
   In the bottom panel, the fitting function is a constant.}
   \label{fig:23}
\end{table}

\begin{figure}[H]
   \centering
   \includegraphics[clip,width=0.45\textwidth]{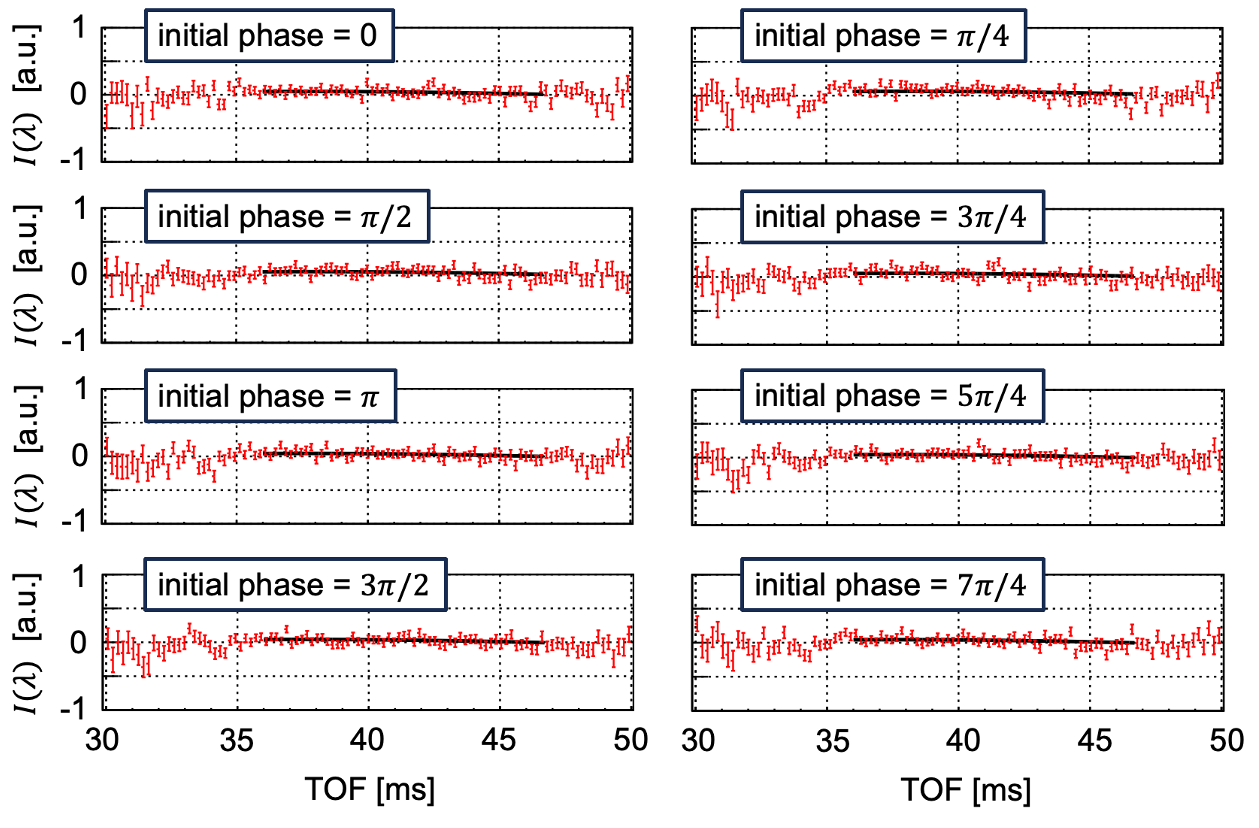}
   \caption{Interference fringes from the sample-in measurements without modulation used for measuring the 250 Hz modulation. In these panels, the red markers represent data points with error bars, and the black curves are the fitting curves. Each result is measured immediately before the modulation of the initial phase, as shown in each panel.}
   \label{fig:24}
\end{figure}

\begin{figure}[H]
   \centering
   \includegraphics[clip,width=0.45\textwidth]{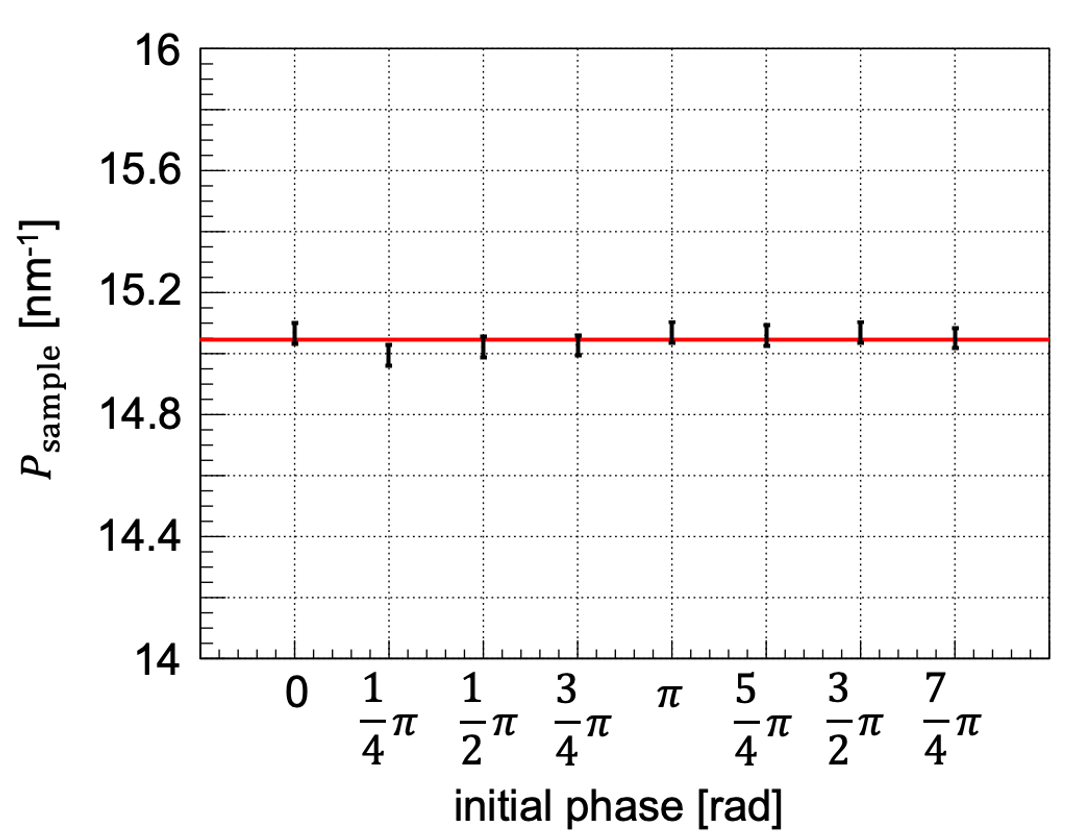}
   \caption{Fitting result of $P_\mathrm{sample}$. The black markers represent data points with error bars, while the red lines show the fitting curves.}
   \label{fig:25}
\end{figure}

\noindent
These results show $P_L = (0.0027\pm0.0041)m + (3.231\pm0.052)$ $\mathrm{nm}$, where $m$ is the measurement number ($m = $1, 2, ..., 8), and $P_\mathrm{IFO} = -10.71 \pm 0.045$ $\mathrm{nm}^{-1}$. $P_L$ is fitted by a linear function with the number of sample-in measurements, $m$, because the drift of the geometric optical length is taken into account. In the subsequent fitting processes for the 250 Hz modulation, these $P_L$ and $P_\mathrm{IFO}$ values are fixed in Eq. \eqref{eq:16}. Next, $P_\mathrm{sample}$ is fitted using a sample-in measurement without modulation. Each sample-in measurement is measured immediately before the modulation of the initial phase to obtain the offset effect from a sample. These results are shown in Figs. \ref{fig:24}-\ref{fig:25}.

\noindent
These results show $P_\mathrm{sample} = 15.05\pm0.012$ $\mathrm{nm}^{-1}$. Since $P_\mathrm{sample}$ is a constant during a series of the sample-in measurements, $P_\mathrm{sample}$ is fitted as a constant. This $P_\mathrm{sample}$ is also fixed in Eq. \eqref{eq:16} for the modulation measurements. Finally, the remaining parameters within the cosine function are fitted using the sample-in measurement with modulation, as shown in Fig. \ref{fig:26}.

\begin{center}
   \includegraphics[clip,width=0.45\textwidth]{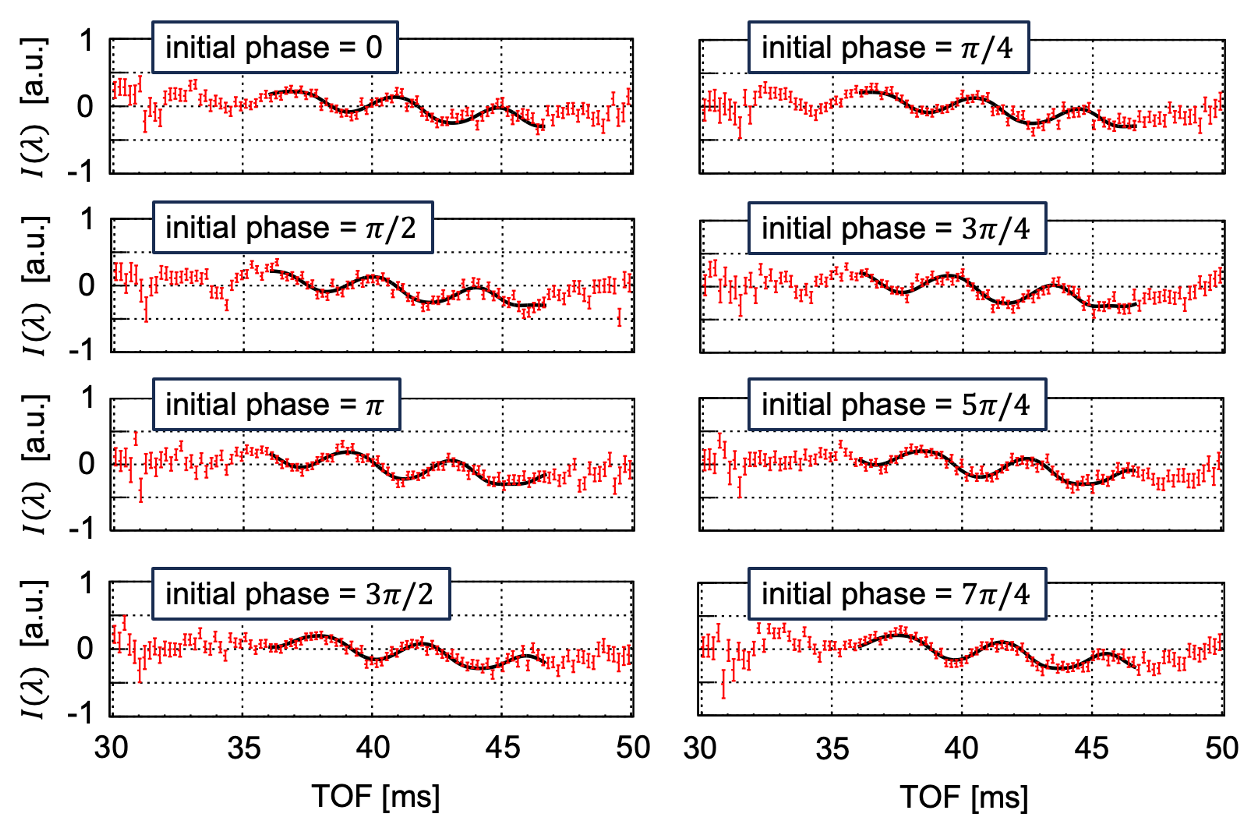}
   \captionof{figure}{Interference fringes from the sample-in measurements with the 250 Hz modulation. In these panels, the red markers represent data points with error bars, while the black curves show the fitting results. Each measurement corresponds to a specific initial modulation phase, as shown in each panel.} % capt-of パッケージを使用
   \label{fig:26}
\end{center}

\begin{figure}[H]
   \centering
   \includegraphics[clip,width=0.45\textwidth]{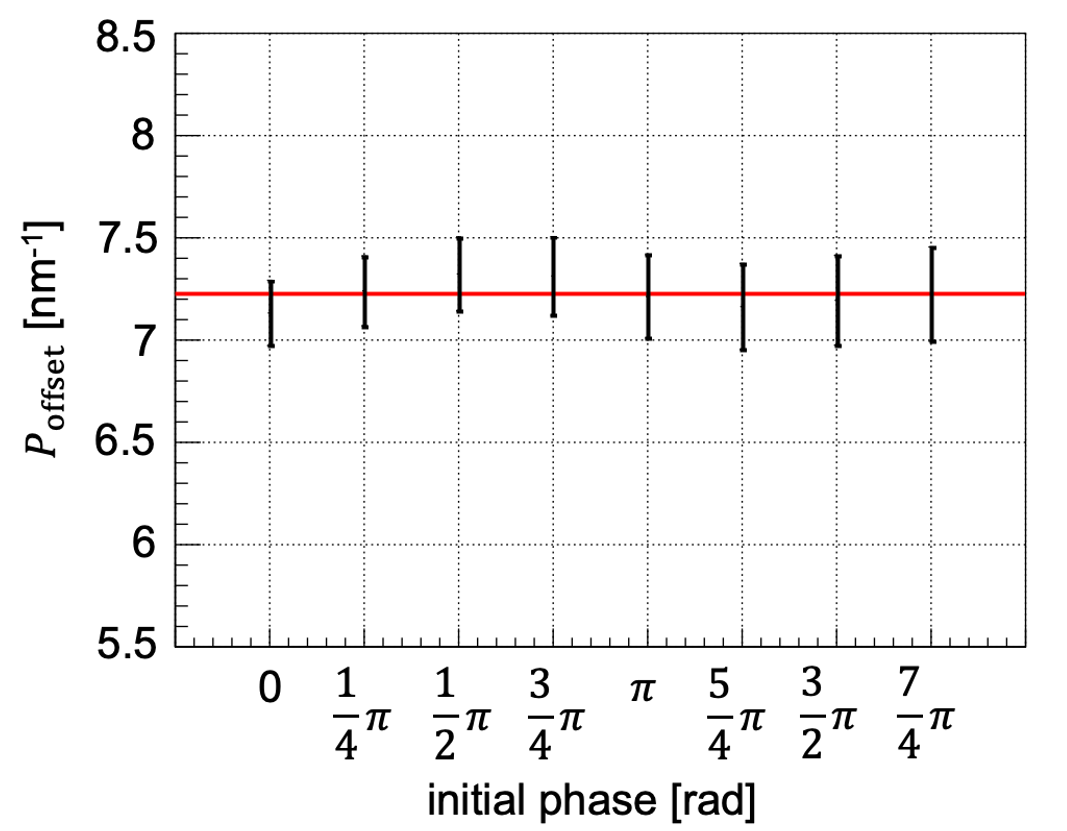}
   \caption{Fitting result of $P_\mathrm{offset}$. The black markers represent data points with error bars, and the red lines indicate the fitting curves.}
   \label{fig:27}
\end{figure}

In this fitting process for the AC parameters, each initial phase $\phi_\mathrm{initial}$ is fixed at 45-degree intervals. This approach assumes that the initial phase error is negligible compared to the frequency error. The values for $P_\mathrm{offset}$, $P_\mathrm{AC}$ and $f_\mathrm{AC}$ values obtained from these sample-in measurements are shown in Figs. \ref{fig:27}-\ref{fig:28}.

% \begin{table}[H]
%    \centering
%    \begin{minipage}{0.235\textwidth}
%        \centering
%        \includegraphics[clip,width=\textwidth]{ACdiff_250Hz_fit.png}
%    \end{minipage}
%    \begin{minipage}{0.23\textwidth}
%        \centering
%        \includegraphics[clip,width=\textwidth]{ACfreq_250Hz_fit.png}
%    \end{minipage}
%    \captionof{figure}{Fitting results of $P_\mathrm{AC}$ in the left panel and $f_\mathrm{AC}$ in the right panel. The black markers represent data points with error bars, and the red lines show the fitting curves.}
%    \label{fig:28}
% \end{table}

\begin{table}[H]
   \centering
   \begin{minipage}{0.45\textwidth}
       \centering
       \includegraphics[clip,width=\textwidth]{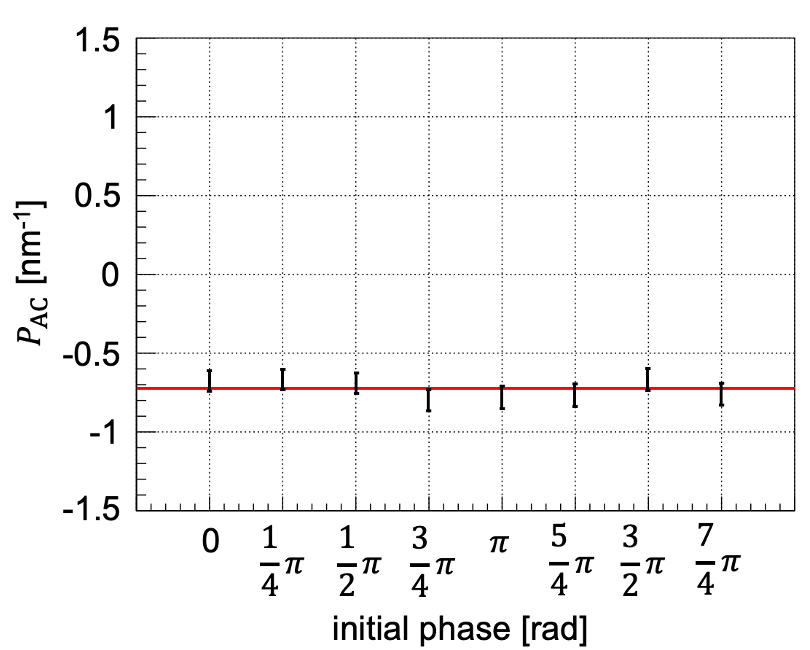}
   \end{minipage}

   \vspace{1em} % 図と図の間隔を調整

   \begin{minipage}{0.45\textwidth}
       \centering
       \includegraphics[clip,width=\textwidth]{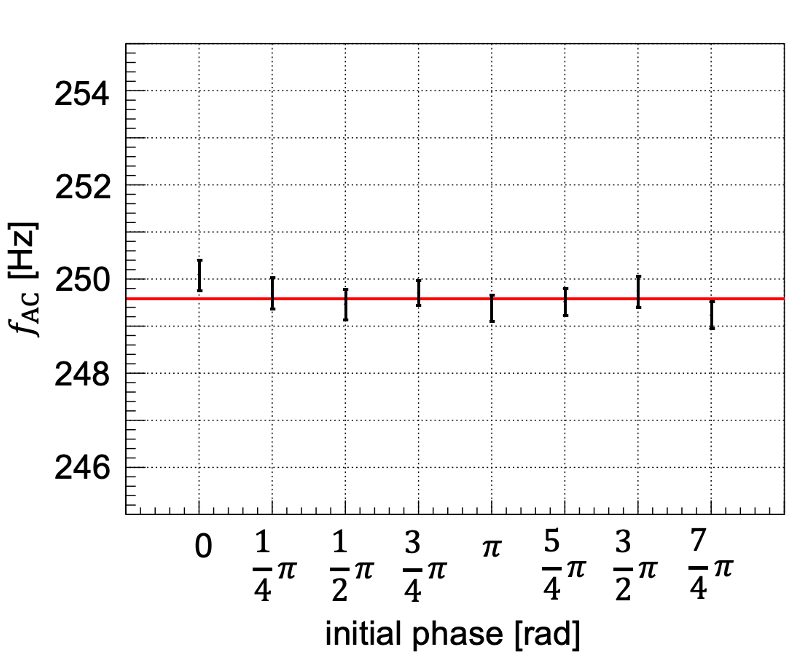}
   \end{minipage}

   \captionof{figure}{Fitting results of $P_\mathrm{AC}$ in the top panel and $f_\mathrm{AC}$ in the bottom panel. 
   The black markers represent data points with error bars, and the red lines show the fitting curves.}
   \label{fig:28}
\end{table}

These results show $P_\mathrm{offset} = 7.225\pm0.068$ $\mathrm{nm}^{-1}$, $P_\mathrm{AC} = -0.7217\pm0.0238$ $\mathrm{nm}^{-1}$, and $f_\mathrm{AC} = 249.59\pm0.11$ Hz.

%\end{comment}

\end{document}